\documentclass[pra,twocolumn]{revtex4}
\usepackage{bbm}
\usepackage{natbib}
\usepackage{graphics}
\usepackage{amsmath}
\usepackage{mathrsfs}
\usepackage{theorem}
\usepackage{float}
\usepackage{hyperref}
\usepackage{rotating}
\usepackage{amssymb}
\usepackage[english]{babel}
\usepackage{color,times}
\usepackage{CJK}
\usepackage{multirow}
\usepackage{lscape}
\usepackage{fancybox}

\newcommand{\ket}[1]{|#1\rangle}

\newcommand{\inp}[2]{\langle #1 | #2\rangle}
\newcommand{\be}{\begin{eqnarray}}
\newcommand{\ee}{\end{eqnarray}}

\begin{document}

\begin{CJK*}{GB}{gbsn}
\title{Edge state, bound state and anomalous dynamics in the Aubry-Andr\'{e}-Haper  system coupled to non-Markovian baths}
\author{H. T. Cui (´Þº£ÌÎ)$^ {a,b}$}
\email{cuiht01335@aliyun.com}
\author{H. Z. Shen (ÉòºêÖ¾)$^ {b}$}
\email{shenhz458@nenu.edu.cn}
\author{M. Qin (ÇØÃ÷)$^{a}$}
\author{X. X. Yi (ÒÂѧϲ)$^{b}$}
\email{yixx@nenu.edu.cn}
\affiliation{$^a$ School of Physics and Optoelectronic Engineering, Ludong University, Yantai 264025
, China}
\affiliation{$^b$ Center for  Quantum Sciences, Northeast Normal University, Changchun 130024, China}
\date{\today}

\begin{abstract}
Bound states and  their influence on the dynamics of an one-dimensional tight-binding system subject to environments are studied in this paper.  We identify  specifically three  kinds of bound states. The first is a discrete bound state (DBS),  of which  the energy level exhibits a  gap from the continuum. The DBS exhibits the similar features of localization as  the  edge states in the system and thus can suppress the decay of system.  The second is a  bound state in the continuum (BIC), which can suppress the system decay too. It is found that the BIC is intimately connected to the edge mode of the system since both of them show almost the same  features of localization and energy. The third one displays a large gap from the continuum and behaves extendible (not localized). Moreover the population of the system on this state  decays partly but not all of them does. This is different from the two former bound states. The time evolution of a single excitation in the system is studied in order to illustrate the influence of the bound states. We found  that both DBS and BIC play an important  role in the time evolution, for example, the excitation becomes localized and not decay depending  on the overlap between the initial state and  the DBS or BIC. Furthermore we observe that the single excitation takes a long-range hopping  when the system falls into the regime of strong localizations. This feature can be understood as the interplay of system localizations  and the  bath-induced long-range correlation.
\end{abstract}

\maketitle
\end{CJK*}

\section{introduction}

In experiments, the environmental effect is unavoidable. A typical example is solid-state quantum devices, which are frequently disturbed by thermal as well as nonthermal environments. This stimulates  the study of open quantum systems. In addition to exploring environmental  effects in quantum devices and shedding light on   the boundary between quantum and classical world, the study on open quantum systems may provide a paradigm to interpret  how an open system  equilibrates  with its surroundings. Especially the localization-delocalization phase transition has been studied intensively in many-body systems with disorders \cite{ai, mbl}, and the quantum many-body scarred state has  been found responsible for the breakdown of thermalization \cite{quantumscar}\cite{thermalization} when there is no disorder in systems.

Recently  bound states that decay exponentially with  small rates have been reported in open systems \cite{john, kofman, bic}. These bound states stem from the shift of system energy levels, induced by the emitted photon that pushes the level beyond the cut-off frequency of the environment \cite{kofman}.  As a result of the appearance of energy gap, the bound states become robust against environment induced decays, and they can prevent quantum systems thermalising  since the  excitations on these states  do not equilibrate. The appearance of bound states is a general feature of  open quantum  systems, independent of the fine structure of the systems. Thus it provides a general way for systems to prevent decoherence.

In fact,  the recent experimental explorations of localization-delocalization transition in cold atomic gas suffer from atom-atom collisions and imperfect trapping \cite{exp-quasidisorder, luschen2017}. The collisions and imperfection can be modeled as environments and  the localized phase would become unstable \cite{luschen2017} due to their influences.  On the theoretical side, it was shown  that  the system exhibits a stretched exponential decay when coupled to a Markovian bath \cite{opendisorder}, then the localization is destroyed and the system is equilibrated finally. Despite these progresses in this direction, the effect of bound states on the dynamics of open system as well as on the localization remains unexplored.

In this paper, we will examine the bound states and the dynamics of an open system. For concreteness,   we consider  a one-dimensional tight-binding  atomic chain with onsite modulation and being coupled to a bosonic bath.  The Hamiltonian of such a  chain is
\be\label{hs}
H_S=\sum_{n=1}^N \left(c^{\dagger}_{n} c_{n+1} +c^{\dagger}_{n+1} c_{n} \right) + \Delta\cos(2\pi \beta n +\phi)c^{\dagger}_n c_{n},\nonumber \\
\ee
where $N$ is the length of atomic chain. $c_n (c^{\dagger}_n)$ is the annihilation (creation) operator of excitation at the $n$-th atomic site.  $\beta$ can be either rational or not, which characterizes two distinct cases. For $\beta=p/q$ with $p$ and $q$ being coprime (commensurate case), the edge mode can occur because of nontrivial topological phase in $H_S $ \cite{lang2012}, which depicts the localization of excitation at boundary. When $\beta$ is a Diophantine number \cite{syj99} (incommensurate case),  $H_S$ corresponds to the Aubry-Andr\'{e}-Haper (AAH) model \cite{aah}, in which a delocalization-localization phase transition happens when $\Delta=2$. Recently it has been demonstrated that AAH model shows the correspondence  to a two-dimensional quantum Hall system \cite{kraus}. Thus the topological edge mode can  be found, in which the excitation would be localized at boundary \cite{kraus}. Moreover AAH model can be realized in cold atomic gas, and the experimental exploration of the  delocalization-localization phase transition has been implemented \cite{exp-quasidisorder}.

The bath and its coupling to the atomic chain are respectively depicted by the following Hamiltonians,
\be
H_B&=& \sum_k \omega_k b_k^{\dagger}b_k;\nonumber\\
H_{int}&=& \sum_{k, n}\left(g_k b_k c_n^{+} + g_k^* b_k^{\dagger} c_n\right)\nonumber,
\ee
where $b_k (b_k^{\dagger})$ is the bosonic annihilation (creation) operator of the $k$-th  mode of bath, and the frequency $\omega_k\geq 0( \forall k)$ consists of a continuum.  $g_k$ characterizes the coupling strength between the lattice site and the $k$-th mode of bath.  We assume that the coupling is so  weak  that  the rotating-wave approximation (RWA) can be applied in $H_{int}$. Then the total Hamiltonian is
\be
H=H_S +H_B  + H_{int}.
\ee
Since there is no particle interaction in $H_S$, the following discussion is restricted to the case of a single excitation, i.e. $\sum_nc^{\dagger}_n c_{n} + \sum_k b_k^{\dagger}b_k=1 $. In  this case the bound state can be determined exactly, and the population dynamics can also be evaluated exactly.  Although the particle interaction is important, we do not try to touch it in the current study since it would make the discussion complicated and ambiguous.

The remainder of this paper is organized as follows. In Sec. II the definition of bound state is presented. Interestingly  a special discrete bound state can be found outside of the continuum $\omega_k$, which does not decay and displays similar localization as the edge mode in $H_S$. However, there also exists a single bound state with very small energy, which is extended and has a certain probability of spontaneous emission. In Sec. III, the population evolution dynamics is calculated,  especially focusing on the interplay of bound state and localization in  system.  It is found that the discrete bound state (DBS) is predominant for the population evolution dynamics. Depending on the overlap of initial state and DBS,  the excitation could become localized against spontaneous emission. Moreover,  the  bound state in the continuum (BIC) can also be identified by finding similar influence on the population evolution dynamics  as the discrete one.   The occurrence of BIC could be attributed to the nontrivial topology in $H_S$ \cite{yang, yao17}.  In Sec.IV  the interplay of  disorder-induced  localization  and bath-induced long-range hopping is studied. We note that the hopping of excitation  can occur over long-range atomic sites, even if the system is  localized strongly. However it is suppressed  greatly when DBS or BIC appears. In Sec.V the long-time behavior of evolution is studied. We observe a very slow decay of excitation in incommensurate case, even if the initial state overlaps with DBS or BIC. However this feature is not found in commensurate case. Finally conclusion is presented in Sec. VI.

\section{Bound state in open systems}

The bound state in open systems is defined  as the discrete energy level of the total Hamilitonian \cite{fain}. As for the continuous spectrum $\omega_k > 0$, the bound state can be determined  only by finding the negative solutions to the Schr\"{o}dinger equation
\be\label{bs}
H\ket{\psi_E}= E \ket{\psi_E}.
\ee
When $E>0$  the solutions could be obtained only for specific $\omega_k$, which thus constitute a continuum. It is the conventional wisdom that the state with frequency inside the continuum  would  leak and radiate out to infinity. However, a bound state in the continuum (BIC) can be found inside the continuum and coexists with extended states, but remains perfectly confined without any radiation\cite{bic}.  Physically the occurrence of BIC can be attributed to the  level resonance  \cite{bic}.  However, it is shown recently that BIC can also be found in  the system with nontrivial topology\cite{yang}.  In order to avoid confusion, we refer to the discrete bound state (DBS)  as the discrete solution to Eq. \eqref{bs}. With respect that BIC can be identified only by the population evolution dynamics, as shown in Appendix C, the following discussion in this section is only suitable for DBS.

For a single excitation, $\ket{\psi_E}$ can be expressed generally as
\be\label{psie}
\ket{\psi_E}&=& \left(\sum_{n=1}^N \alpha_n \ket{1}_n\ket{0}^{\otimes (N-1)} \right)\otimes \ket{0}^{\otimes M} + \nonumber \\ &&\ket{0}^{\otimes N} \otimes \left(\sum_{k=1}^{M} \beta_k \ket{1}_k\ket{0}^{\otimes (M-1)} \right),
\ee
where $\ket{1}_n = c_n^{\dagger}\ket{0}_n$ denotes the occupation of the $n$-th lattice site, $\ket{0}_k$ is the vacuum state of $b_k$ and $\ket{1}_k=b_k^{\dagger}\ket{0}_k$, and $M$ denotes the number of modes of bath. Substituting Eq.  (\ref{psie}) into Eq.  (\ref{bs}), one obtains
\addtocounter{equation}{1}
\begin{align}\label{bsea}
&\left(\alpha_{n+1} + \alpha_{n-1}\right) +\Delta \cos(2\pi \beta n +\phi)\alpha_n + \sum_{k=1}^{M} g_k \beta_k = E \alpha_n;  \tag{\theequation a} \\
&\omega_k \beta_k + g_k^*\sum_{n=1}^N \alpha_n= E \beta_k. \tag{\theequation b}\label{bseb}
\end{align}
According to Eq.  (\ref{bseb}),
\be
\beta_k=\frac{ g_k^*}{E-\omega_k}\sum_{n=1}^N \alpha_n.
\ee
Substitute the expression of $\beta_k$ into Eq. (\ref{bsea}), and then
\be
\left(\alpha_{n+1} + \alpha_{n-1}\right) &+&\Delta \cos(2\pi \beta n +\phi)\alpha_n +\nonumber \\
&& \left(\sum_{k=1}^{M} \frac{ \left|g_k\right|^2}{E-\omega_k}\right) \sum_{n=1}^N \alpha_n= E \alpha_n.\nonumber
\ee
As for the continuous spectrum $\omega_k$,
\be\label{int}
\sum_{k=1}^{M} \frac{ \left|g_k\right|^2}{E-\omega_k} \rightarrow \int_0^{\infty} \frac{J(\omega)}{E-\omega} \text{d}\omega,
\ee
where the spectral density $J(\omega)= \sum_{k=1}^{M}  \left|g_k\right|^2 \delta\left(\omega-\omega_k\right)$. Then one has
\be\label{boundeqn}
\left(\alpha_{n+1} + \alpha_{n-1}\right) &+& \Delta \cos(2\pi \beta n +\phi)\alpha_n  +\nonumber \\
&& \int_0^{\infty} \text{d}\omega\frac{J(\omega)}{E-\omega} \sum_{n=1}^N \alpha_n = E \alpha_n.
\ee
With respect that  the integrals Eq.  (\ref{int}) is divergent for $E>0$,  the solutions to Eq. \eqref{boundeqn} can  be acquired  only for $E<0$.  Physically  the last term at the left hand of  Eq. \eqref{boundeqn}  characterizes a homogenous hopping of excitation over atomic sites.  As will be displayed in Sec. IV, the interplay of this effective long-range correlation and localization in system will impose a significant effect on the population evolution dynamics.

For concreteness, the  spectral function  is chosen as
\be\label{j}
J(\omega)= \eta \omega \left(\frac{\omega}{\omega_c}\right)^{s-1}e^{-\omega/\omega_c},
\ee
where $\eta$ characterizes the coupling strength between system and bath. The bath can be classified  as sub-Ohmic ($s<1$), Ohmic ($s=1$) and super-Ohmic ($s>1$) \cite{leggett}. Physically  Eq. \eqref{j} characterizes the damping movement of electrons in a potential, and thus provides a general picture for the dissipation of excitation in system. When disorder exists, it is expected that the competition between localization and the bath-induced dissipation  would have a major influence on the dynamics of excitation.  So the choice for  $J(\omega)$ is suitable for the current interest. As for $s$, it is shown in Appendix A that  the  discrete solutions to Eq.  \eqref{boundeqn} show negligible dependence on the value of $s$, except for the ground state. Thus the following  discussion is restricted  to the case of $s=1$.  $\omega_c$ is the cutoff frequency of the bath spectrum, beyond  which the spectral density starts to fall off. Hence, it determines a regime of frequency in bath, which is predominant for dissipation. In general the value of $\omega_c$ depends on specific environment. However as shown in  Appendix A, $\omega_c$ shows a negligible effect on the solutions to Eq.  \eqref{boundeqn}, except for the ground state. Hence  $\omega_c=10$ is chosen in order to ensure $\Delta/\omega_c <1$ \cite{leggett}. In addition, an exceptional case can be found for the minimal solution $E_0$,  which exhibits heavy dependence on the size of system and the properties of the bath. Thus the level $E_0$ would show distinct behavior.

Eq. \eqref{boundeqn} constitutes a linear system of equations for variable $\alpha_n$. The values of  $E$ can be determined  by  finding out the zero points of determinant of coefficient matrix. However, noting that  $E$ is also involved in the integrals, one thus  has to appeal to numerics. Our evaluation shows that there are $N$ negative solutions to $E$ at most. Consequently as for large $N$, these solutions could  constitute a band. Actually we find that the band is significantly overlapped with that in $H_S$ for $E\leq 0$. This feature can be attributed to the weak system-bath couplings: The bath cannot provide enough energy for the transition between different bands. It is difficult to determine the continuous spectrum $E$ in numerics. As a consequence we try to  find the discrete $E$ in band gap, which is more tractable in numerics and meaningful in physics.  Moreover it is expected that the discrete solution would be related intimately with the edge model in $H_S$ and thus could be stable against decoherence.  So the remaining discussion in this section would  focus on the discrete solutions occurring in gap instead. The terminology of DBS is designated as the special solution in this place. For this purpose, two situations are discussed respectively: commensurate ($\beta=1/3$) and incommensurate ($\beta=\left(1 + \sqrt{5}\right)/2$) cases, in which DBS behaves differently.

\begin{figure}[t]
\center
\includegraphics[width=8cm]{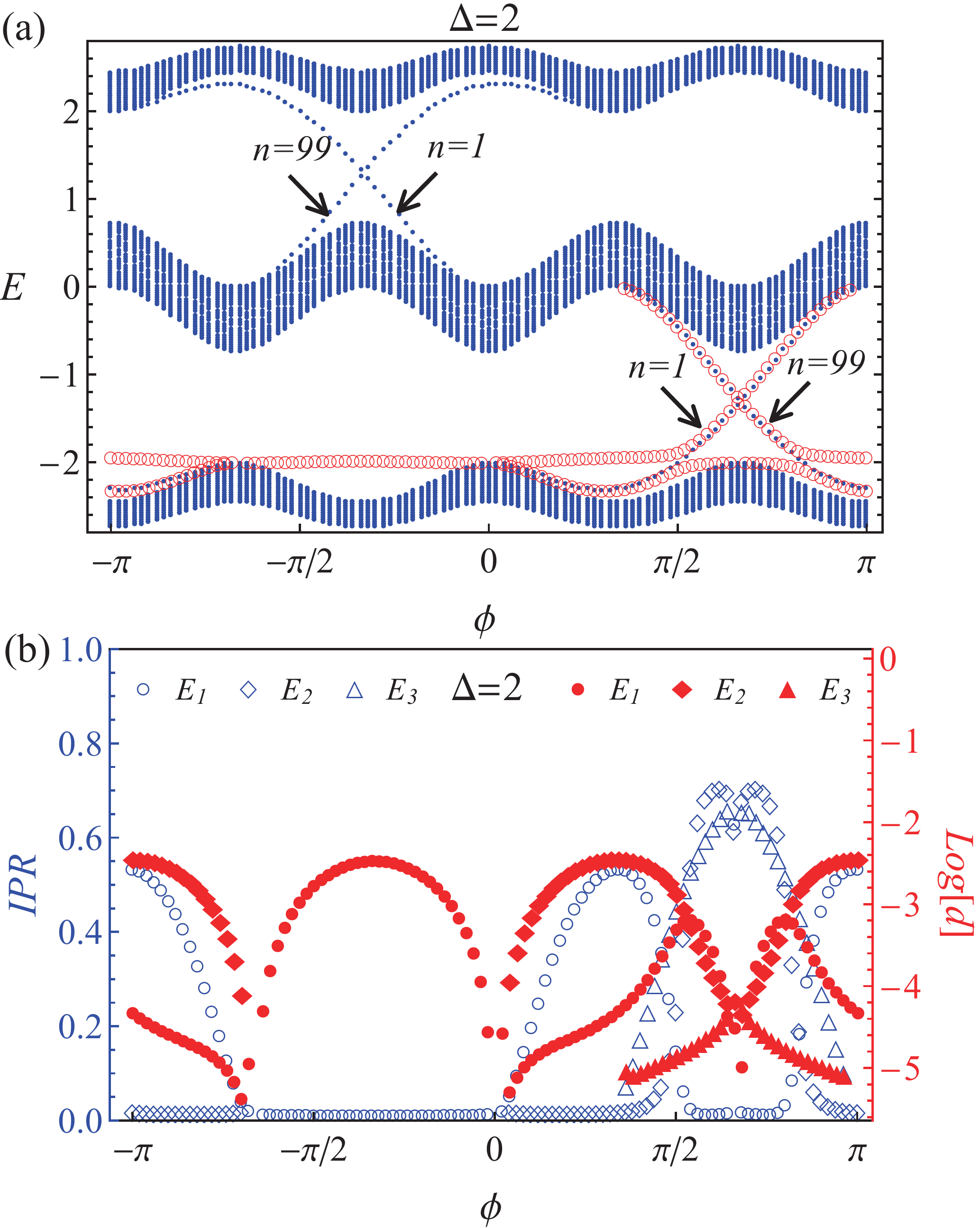}
\caption{(Color online)(a)Plots of the energy levels for $H_S$ with $\beta=1/3$  and $\Delta=2$  (blue  point) and the discrete bound states (red empty circle)  when $E<0$. The labels $n=1 (99)$ denote  the site, at which excitation is localized; (b) The plots of IPR (in blue empty circle) and  $d$ (in red solid triangle) for DBS in the panel (a).  The labels $E_1$, $E_2$ and $E_3$ denote the levels of DBS by increscent order. $N=99$,  $s=1$, $\eta=0.1$ and $\omega_c=10$ are chosen for all plots.}
\label{fig:com-be}
\end{figure}

\subsection{Commensurate case: $\beta=1/3$}

When $\beta=p/q$ ($p$ and $q$ being coprime), the spectrum of $H_S$ consists of $q$  bands. As an exemplification, the spectrum of $H_S$ are demonstrated  for $\beta=1/3$ $\Delta=2$ under open boundary in Fig. \ref{fig:com-be}(a) (solid points). The edge mode, plotted by the discrete solid points in gap, depicts the localization of excitation at ends. In contrast the state in band is extended.  By solving Eq. \eqref{boundeqn} three discrete solutions at most can be found in  gap when $E<0$,  which are highlighted by red empty circles in Fig.\ref{fig:com-be}(a). It is evident that two different features can be observed for these solutions.  One is  the DBS that has nearly the same  energy as the edge mode in $H_S$. We find that it exhibits similar localization as the  edge state, and thus could be considered as the renormalization of edge state.  The other is the DBS that has different energy from the edge mode. We find that  it is extended instead,  as shown by  the inverse participation ratio (IPR)  $\text{IPR}=\sum_n \left|\alpha_n\right|^4$  in Fig. \ref{fig:com-be}(b), and thus comes from  the transition of the state in band.

The unnormalized probability of spontaneous emission  defined as
\be
d= \sum_k \left|\beta_k\right|^2= \left|\sum_{n=1}^N\alpha_n\right|^2\int_0^{\infty} \frac{J(\omega)}{(E-\omega)^2} \text{d}\omega,
\ee
is calculated for all DBSs, as shown in Fig.\ref{fig:com-be}(b) by $\log d$. It is clear that $d$ has an amplitude not larger than $\sim 10^{-2}$. This picture means that DBS is robust against spontaneous emission.

However a single special solution  $E_0 \sim 23.13$ to Eq. \eqref{boundeqn} can be found, for which the corresponding $\text{IPR}\sim1/99\approx 0.01$ and the probability of spontaneous emission is $\sim 0.405$. Furthermore we also find that  $E_0$ is almost independent of $\phi$ and $\Delta$. For example,  $E_0\sim -23.13$ for $\Delta=1$ and $-23.356< E_0<-23.31$ for $\Delta=4$. Instead it shows significant dependence on the system size $N$ and the properties of bath, as shown in Appendix A. It thus means that this special bound state is extended, and characterizes strong entanglement between the system and bath.

\begin{figure*}
\center
\includegraphics[width=15cm]{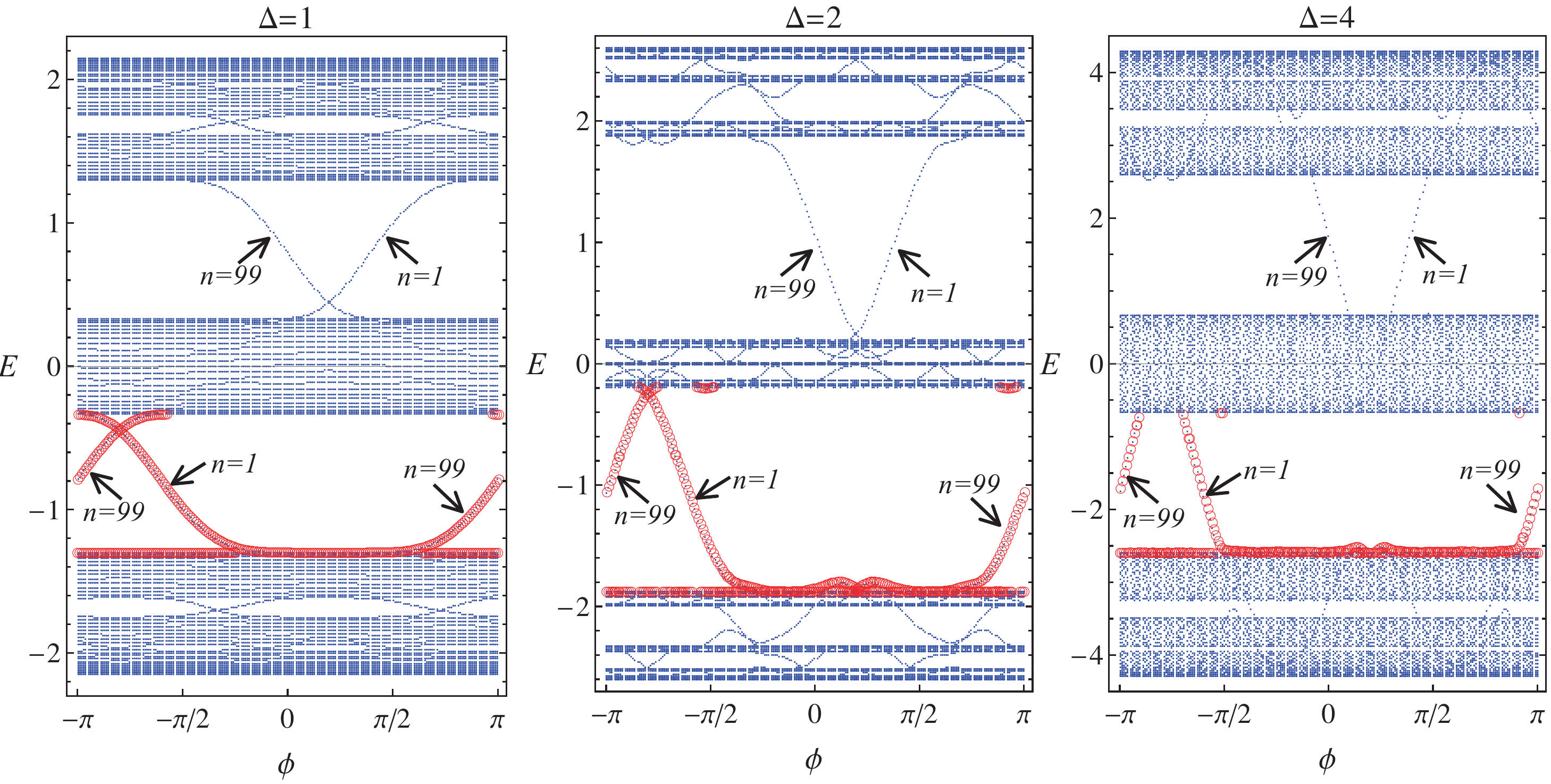}
\caption{(Color online) Plots of the levels (blue solid point) of $H_S$ with $\beta= \left(1 + \sqrt{5}\right)/2$ for (a) $\Delta=1$, (b) $\Delta=2$, (c) $\Delta=4$ and DBS for $E<0$ (red empty-circle). The other parameters are same to those in Fig. \ref{fig:com-be}. The label $n=1 (99)$ denotes the site, occupied by  excitation in edge mode and  DBS.}
\label{fig:incom-be}
\end{figure*}

\subsection{Incommensurate case: $\beta= \left(1 + \sqrt{5}\right)/2$}

The localization-delocalization phase transition can occur  when $\beta$ is a Diophantine number \cite{syj99}. With respect that the Diophantine number can be approached infinitely by rational numbers, the system is actually quasi-periodic, which induces a fractal structure in band as shown in Fig.\ref{fig:incom-be}. Furthermore there is a critical point  $\Delta =2$ in $H_S$,  which separates the delocalized phase ($\Delta<2$) from the localized phase ($\Delta>2$). In the delocalized phase all eigenstates tend to be extended. In contrast they show strong localization  in localized phase. The in-gap edge state can also be found under open boundary condition since $H_S$ is  equivalent to a two-dimensional Hofstadter model \cite{kraus}.

As for concreteness, $\beta= \left(1 + \sqrt{5}\right)/2$ is chosen. By solving Eq. \eqref{boundeqn}, DBS can be decided exactly, which is highlighted by red empty circles in Fig. \ref{fig:incom-be} for $\Delta=1, 2, 4$ respectively. It is evident that there are two main gaps as well as several mini gaps. Although  some discrete solutions  may be found in the mini gaps, the following discussion will focus on the solutions in the two main energy gaps since the fractal bands are meaningless in physics. It should be pointed out that we do not try to discuss the variance of critical point because of the coupling to the bath. So the following discussion for $\Delta=2$ is just to show the influence of  quasi-disorder.

An interesting  feature in this case is  that  the discrete solution in main gap  shows an apparent correspondence to the edge mode. This phenomenon could be attributed to the robustness of quasi-disorder against dissipation. So there is no transition occurring for the state in band, and  the edge mode is renormalized as the DBS. In addition the localization in DBS is enhanced with the increment of $\Delta$,  as shown by IPR in  Fig.\ref{fig:incom-bs} in Appendix B. The corresponding  $d$  also tends to be disappearing, which implies that the spontaneous emission of excitation is suppressed greatly.

Similar to the commensurate case,  a single special  solution  $E_0$ can also be found. For instance we find that $E_0\sim -23.13, -23.17$ for $\Delta=1, 2$ and $\sim -23.351 <E_0 < \sim-25.32 $  for $\Delta=4$. Moreover the corresponding $\text{IPR}\approx 0.01$ and $d \approx 0.4$, independent of $\Delta$ and $\phi$.

\subsection{Further  Discussion}

In conclusion the DBS can always be found in gap, which is connected intimately with the edge mode in $H_S$.  A common property for DBS is the disappearing spontaneous emission, and thus the excitation can be preserved in  system against decoherence. While the DBS shows one-to-one correspondence to the edge mode in the incommmensurate case,   an additional DBS can be found in commensurate case,  which has distinct energy from  the edge mode and behaves extended instead, as shown in Fig. \ref{fig:com-be}(a).  This phenomena can be attributed to the quasi-disorder in $H_S$, which makes the system stable against the transition induced by the coupling to a bath. In addition, we also find that the corresponding IPR is smaller than 1. The reason is  the competition between the disorder-induced localization  and the coupling-induced long-range correlation that  makes the excitation hop in different sites. A detailed discussion for IPR can be found  in Appendix B.

Another common picture is the existence of a special bound state $E_0$, which is extended and has a  probability of spontaneous emission  $\sim 0.4$. Moreover this special state exhibits strong dependence on the properties of bath and the system size $N$. Consequently  $E_0$ characterizes the equilibrium between localization and dissipation, and thus be useless for the storage of quantum information.

\section{time evolution}

The population evolution of single excitation in system is discussed in this section, in order to demonstrate the strong influence of bound state. The evolution equation is written as
\be\label{evolution}
\mathbbm{i}\frac{\partial }{\partial t}\alpha_n(t)&=& \left[\alpha_{n+1}(t) + \alpha_{n-1}(t)\right]+ \Delta \cos(2\pi \beta n +\phi)\alpha_n(t) \nonumber\\
&&- \mathbbm{i} \sum_{n=1}^N \int_0^t \text{d}\tau\alpha_n(\tau)f(t-\tau),
\ee
where $\mathbbm{i}$ is the imaginary unit, and the memory kernel  $f(t-\tau)= \frac{\eta}{\omega_c^{s-1}} \frac{\Gamma(s+1)}{\left[\mathbbm{i}(t-\tau) + 1/\omega_c\right]^{s+1}}$ is responsible for dissipation.  Because of involved integrals,  numerical evaluation has to be implemented to find out $\alpha_n(t)$. Our way is to rewrite  the integrals as a summation with  suitable step length. Then by solving Eq.  (\ref{evolution}) iteratively, $\alpha_n(t)$ can be determined finally. %Although this way limits the evaluated length of time and the size of system, the crucial feature of evolution can be disclosed.

Formally when the bound state occurs,  $\ket{\psi(t)}$ can be decomposed  into two parts,  i.e.
\be\label{psit}
\ket{\psi(t)}= \sum \alpha_b\ket{\psi_b} e^{-\mathbbm{i}E_b t} + \int\text{d}E_c  \alpha(E_c)e^{-\mathbbm{i}E_c t } \ket{\psi_c}.
\ee
The summation is over all bound states $\ket{\psi_b}$ with  energy $E_b$, which means unitary evolution and thus is responsible for the robustness of excitation.  While the integrals over the continuum  $E_c$  is responsible for the decay of excitation, which tends to be vanish after a long time. As a result the bound states will determine completely the final state of system. In order to highlight the effect of DBS or BIC, we choose the initial state $\ket{\psi(t=0)}=\sum_n \alpha_n(0)\ket{n}$ with a single excitation located  at atomic site $n_0=1$ and $n_0=99$ respectively. The corresponding revival probability of excitation $\left|\alpha_1(t)\right|^2$ and $\left|\alpha_{99}(t)\right|^2$  are calculated, as well as the corresponding $\text{IPR}_{1(99)}$.  Three distinct behaviors can be found for the population evolution of single excitation. First the excitation becomes localized at its initial site.  Second the excitation can hop to a different site from its initial one. Thirdly the evolution is dissipative and excitation could be absorbed finally by bath. %We will show that these three pictures are intimately connected with the DBS or BIC.

\begin{figure}
\center
\includegraphics[width=8cm]{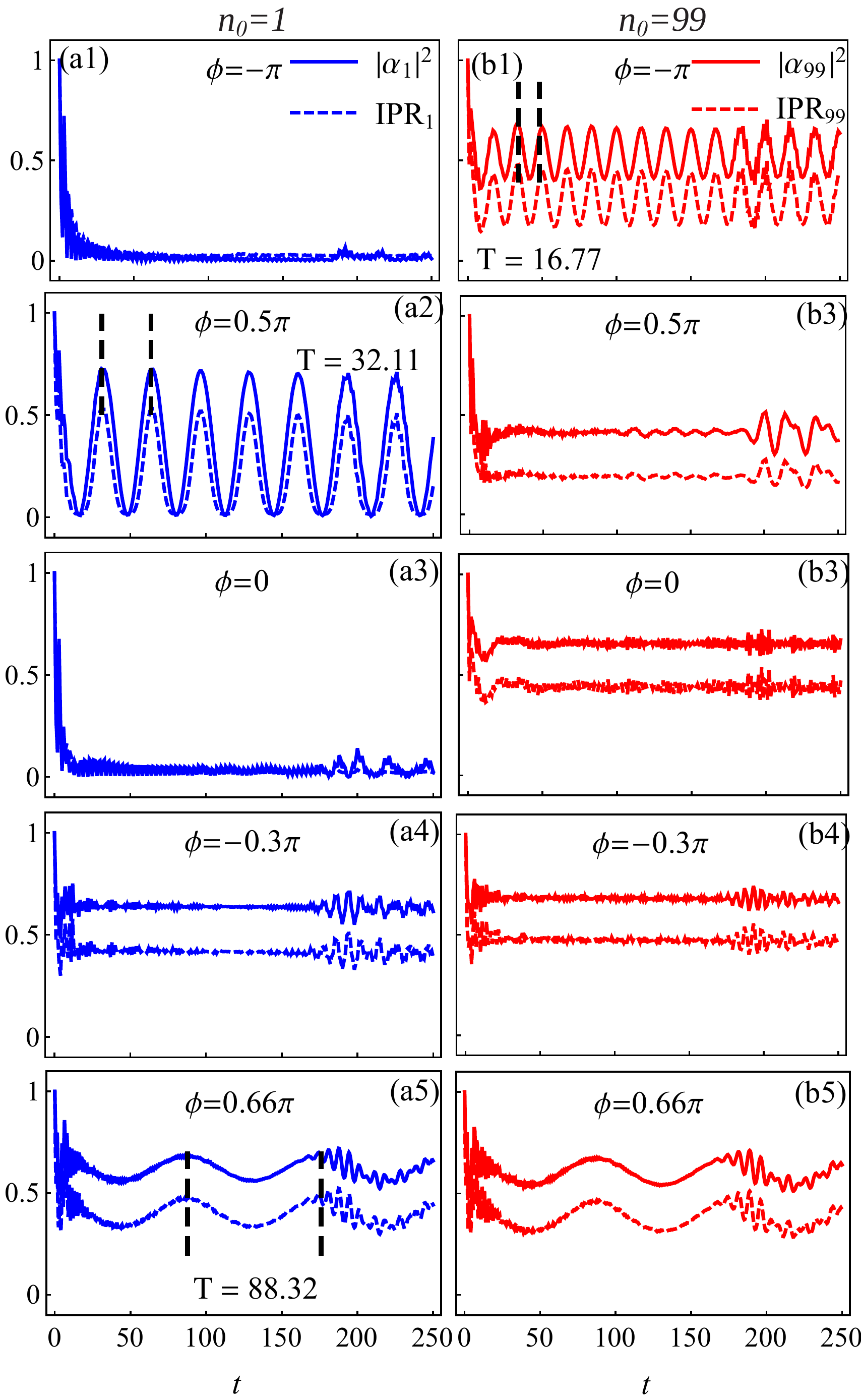}
\caption{(Color online) The evolution of survival probability $\left|\alpha_n\right|^2 (n=1,99)$ (solid line) and the corresponding $\text{IPR}_n$ (dashed line) for a single excitation initially at $n_0=1$ (left column) or $n_0=99$ (right column).  $\beta=1/3$ and $\Delta=2$ are chosen, and the other parameters  are same to those in Fig.\ref{fig:com-be}. }
\label{fig:com-evolution}
\end{figure}

\begin{figure}
\center
\includegraphics[width=6cm]{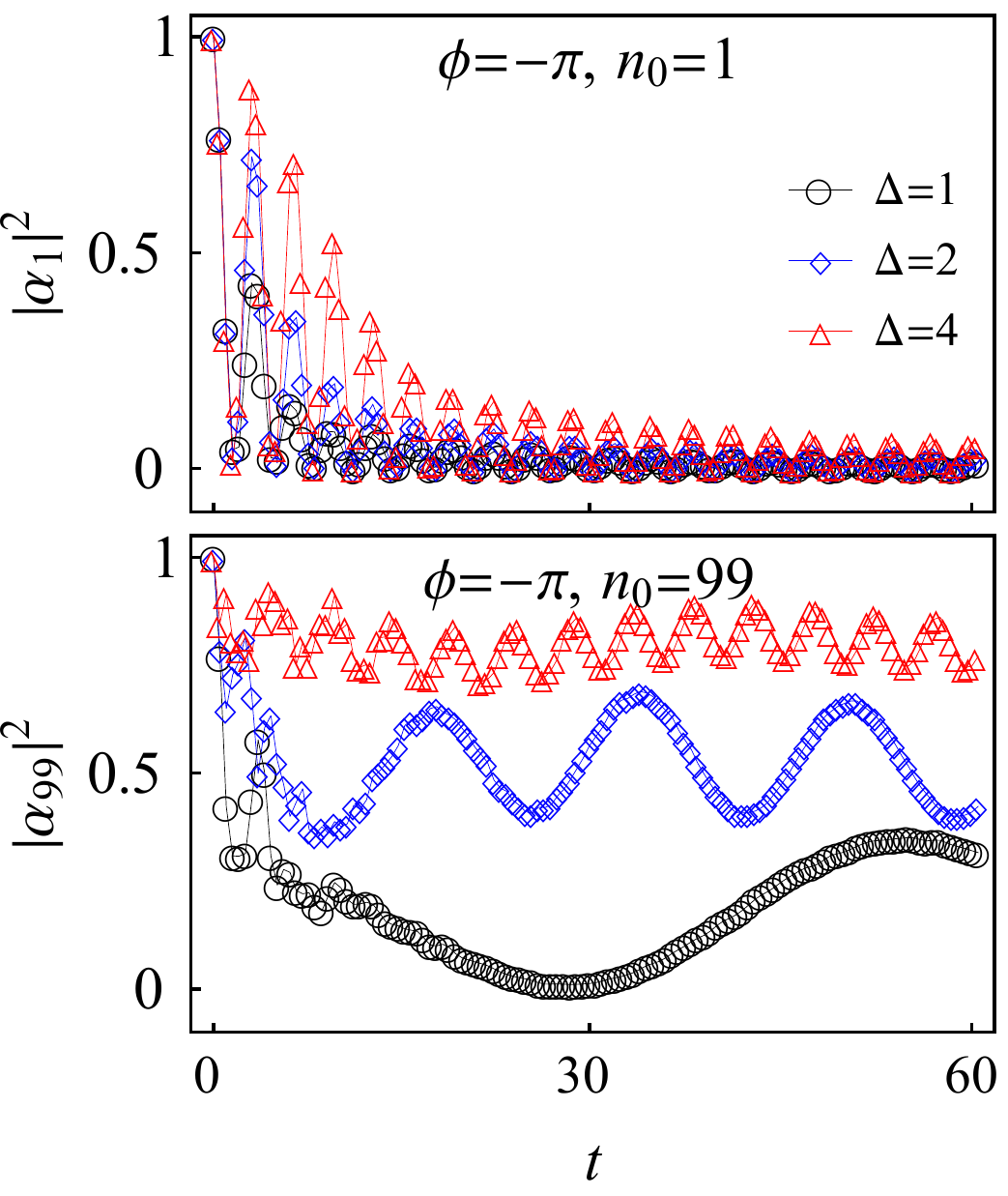}
\caption{(Color online) The plots of survival probability for different $\Delta$ when $\beta=1/3$ and $\phi=-\pi$.  The other parameters are same to those in Fig. \ref{fig:com-evolution}. }
\label{fig:com-delta}
\end{figure}

\subsection{Commensurate case: $\beta=1/3$}

Five different cases are plotted in Fig. \ref{fig:com-evolution}. For $\phi=-\pi$ two DBSs can be found  when  $E<0$,  as shown in Fig. \ref{fig:com-be}(a); One is overlapped with the edge state and shows strong localization at site $n=99$.  Whereas the other is extended. It is clear that the survival probability $\left|\alpha_{99}\right|^2$ shows a stable  oscillation around  0.5 for excitation located initially at site $n_0=99$, as shown in  Fig.\ref{fig:com-evolution}(b1). This oscillation stems from the interference of two DBSs, that can be affirmed by measuring the frequency of oscillation. As shown in  Fig.\ref{fig:com-evolution}(b1), the period of oscillation is $T=16.77$. Then the frequency $\omega=2\pi/T=0.3747$, which is closed to the energy difference $\delta E= 0.3768$ of the two DBSs. The slight difference comes from the computational error. However, $\left|\alpha_{1}\right|^2$  for $n_0=1$ displays a rapid decay, as shown in  Fig.\ref{fig:com-evolution}(a1). The same features can also be found for IPR (dashed line in Fig.\ref{fig:com-evolution}).  The observation implies that DBS would determine completely the population evolution: When the initial state is overlapped with DBS, the excitation can be preserved with a large probability. While if not, the information of initial state would be erased completely. So in this sense the edge state would be  renormalized as a DBS. It should be pointed out that the weak fluctuation of survival probability for $t >\sim 180$ comes from the accumulation of computational error in solving Eq. \eqref{evolution} iteratively.

%The similar behavior can also be found for $\phi=0.3\pi$,  in which there is two DBS in gap. The stable oscillation of $\left|\alpha_{1}\right|^2$ and $\text{IPR}_{1}$, shown in Fig.\ref{fig:com-evolution}(a2), also comes from the interference of two DBSs.

Similar phenomena can also be observed  for  $\phi=0.5\pi$, in which there are three DBSs, as shown in Fig. \ref{fig:com-be}(a). Two of them show  similar localization as the edge states. The third behaves  extended instead.  It is noted that $\left|\alpha_{1}\right|^2$ shows a stable oscillation with period  $T=32.11$ because of  the interference of the two lowest DBSs, with the energy difference $\delta E= 0.1953$. At the same time $\left|\alpha_{99}\right|^2$ becomes stable when the initial state is overlapped with the DBS, which shows strong localization at $n=99$. An interesting situation is $\phi=0.66\pi$: There are two DBSs with  localizations at  $n=1$ and $n=99$ respectively. They are closed to each other in energy  as shown in Fig.\ref{fig:com-be}(a). Consequently a stable oscillation can be found for both $\left|\alpha_{1}\right|^2$ and $\left|\alpha_{99}\right|^2$  because of the interference, as shown in Fig.\ref{fig:com-be}(a5) and (b5).   As will be discussed  in next section, this interference  induces an end-to-end hopping of excitation.

A special case happens  for $\phi=0$, in which there is no DBS when $E<0$. In contrast to the rapid decay of  $\left|\alpha_{1}\right|^2$, a stable evolution can be noted for the excitation initially located at $n_0=99$, as shown in Fig.\ref{fig:com-be}(a3) and (b3). This phenomenon can be attributed to the occurrence of  BIC \cite{bic},  as shown  in Appendix C. Generally  BIC is induced by the level resonance \cite{bic}. However in the present discussion BIC could be understood by the nontrivial topology in $H_S$ \cite{yang, yao17}.  It is clear that both DBS and BIC manifest similar influence on the population evolution dynamics. Another  exemplification of BIC can be found when $\phi=-0.3\pi$. Under this circumstance, there are  two edge states in $H_S$ when $E>0$ with the localization at $n=1$ and $n=99$ respectively.  Consequently both $\left|\alpha_{1}\right|^2$ and $\left|\alpha_{99}\right|^2$ show stable evolution, as shown in   Fig.\ref{fig:com-evolution} (a4) and (b4).

The localization is enhanced with the increment of  $\Delta$, as shown by $\left|\alpha_{99}\right|^2$ in Fig. \ref{fig:com-delta} for $\phi=-\pi$. At the same time the decay of $\left|\alpha_{1}\right|^2$ also becomes stretched slightly. This feature can be attributed to the trapping effect of on-site potential.

\begin{figure}[t]
\center
\includegraphics[width=8.5cm]{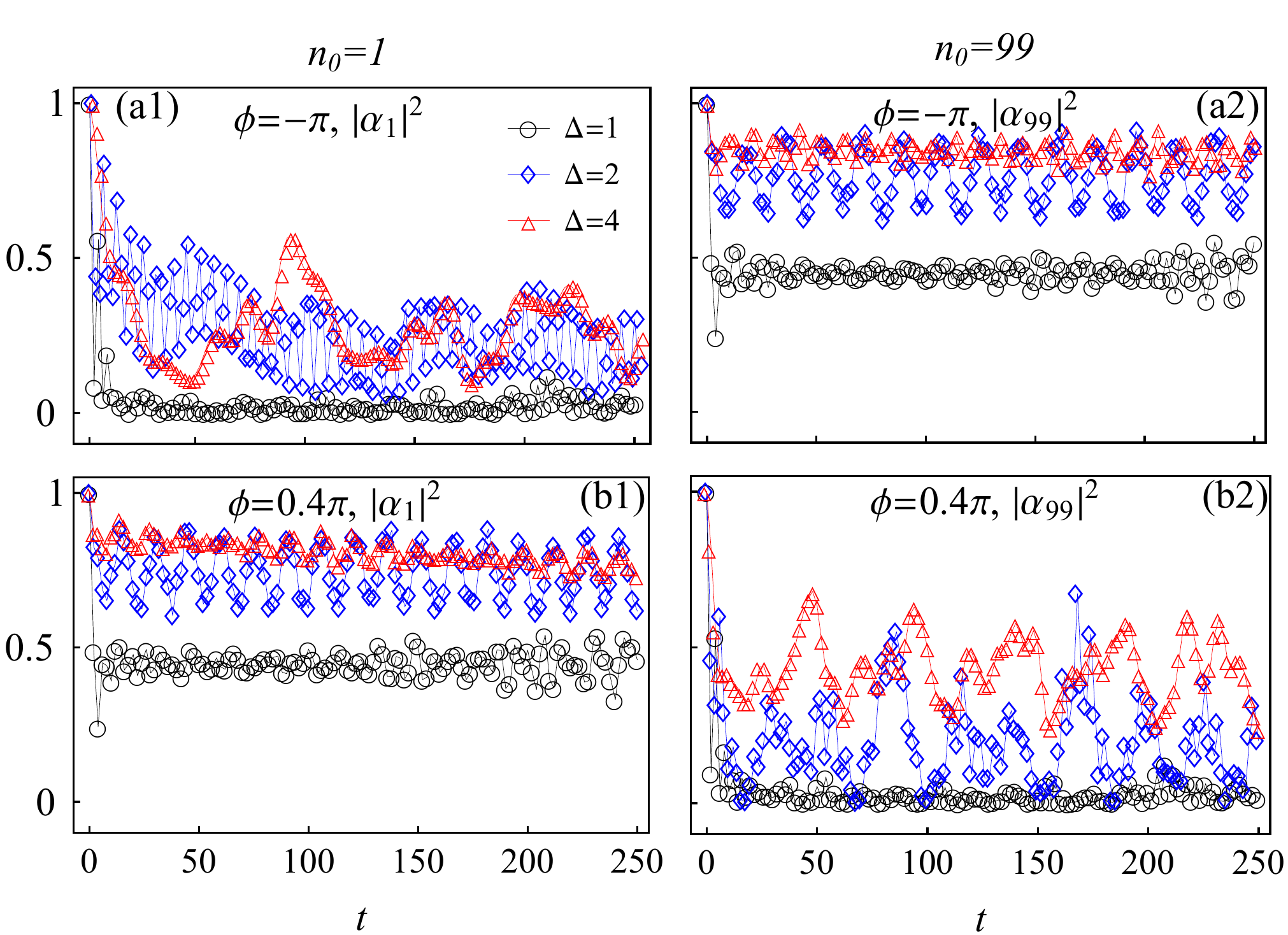}
\caption{(Color online) The plots of survival probability $\left|\alpha_{1(99)}\right|^2$  for excitation initially at site $n_0=1$ (left column) and $n_0=99$ (right column) versus $t$ when $\Delta=1, 2,4$.  $\beta=\left(1 + \sqrt{5}\right)/2$ are chosen for these plots. The other parameters are same to those in Fig. \ref{fig:com-evolution} }
\label{fig:incom-evolution}
\end{figure}

\subsection{Incommensurate: $\beta= \left(1 + \sqrt{5}\right)/2$ }

Two distinct phases  can be identified in this case: delocalized phase ($\Delta<2$), in which the system is extendible, and the localized phase ($\Delta>2$), in which the system displays strong localization. As exemplifications, the cases of $\phi=-\pi$ and $\phi=0.4\pi$ are studied in details, for which there is a DBS and a BIC with localization at site $n=99$ and $1$ respectively,  as shown in Fig. \ref{fig:incom-be}. As expected, the stable evolution can be found for excitation located initially at $n_0=99$ or $n_0=1$,  as shown in Fig. \ref{fig:incom-evolution} (a2) and (b1). Furthermore we also note that although the survival probability is enhanced with the increment of $\Delta$, a strange feature can be found in Fig. \ref{fig:incom-evolution}(b1), where   $\left|\alpha_{1}\right|^2$ declines smoothly when $\Delta=4$. This abnormal feature will be discussed alone in Sec. V.

However The picture becomes different  when the initial state is not overlapped with any DBS or BIC. For example, the survival probability $\left|\alpha_{1}\right|^2$ for $\phi=-\pi$ exhibits  a rapid decay when $\Delta=1$. However when $\Delta=2, 4$, a significant recurrence can  be found  for $\left|\alpha_{1}\right|^2$, as shown in  Fig. \ref{fig:incom-evolution} (a1). This  feature could be attributed to the influence of the bound states other than DBS and BIC. As stated in Sec. II, the solutions to Eq. \eqref{boundeqn} other than the discrete ones in gap, constitute the band, which become more localized with the increment of $\Delta$. Consequently when the initial state is  overlapped substantially  with the states in band, the interference of states thus  would induce the temporal revival of  $\left|\alpha_{1}\right|^2$. This explanation can be verified in further by noting that the recurrence is absent in commensurate case and for $\Delta=1$, in which the states in band are extended or delocalized. Similar picture can also be found for $\left|\alpha_{99}\right|^2$ when $\phi=0.4\pi$, as shown in Fig. \ref{fig:incom-evolution} (b2).

\subsection{Further Discussion}

It is evident that the bound state  is predominant in the population evolution. Dependent on the overlap of initial state and DBS or BIC, the survival probability of excitation can become stable against dissipation. For both commensurate  and incommensurate cases, the excitation can be preserved in system with a large probability if the initial state is overlapped with DBS or BIC. In contrast if not, two different features would be obtained in our discussion. When $H_S$ is  commensurate or in delocalized phase ($\Delta<2$), the population evolution  is dissipative. However in the localized phase of $H_S$ ($\Delta>2$), it can show a recurrence due to the strong localization of $H_S$, which cannot be destroyed completely by coupling to a bath.

An interesting question is the excitation dynamics when there is no DBS or BIC.  As shown  by the integrals in \eqref{boundeqn}, an effective long-range correlation in atomic sites is inspired by the coupling to bath, which is responsible for the dissipation of excitation. However the quasi-disorder in $H_S$ tends to localize the excitation in the system. Hence it is expected that the interplay of the long-range correlation and the localization induced by quasi-disorder would inspire exotic dynamics of excitation. In the next section, we shed light on the influence of this interplay.

\begin{figure}
\center
\includegraphics[width=8.5cm]{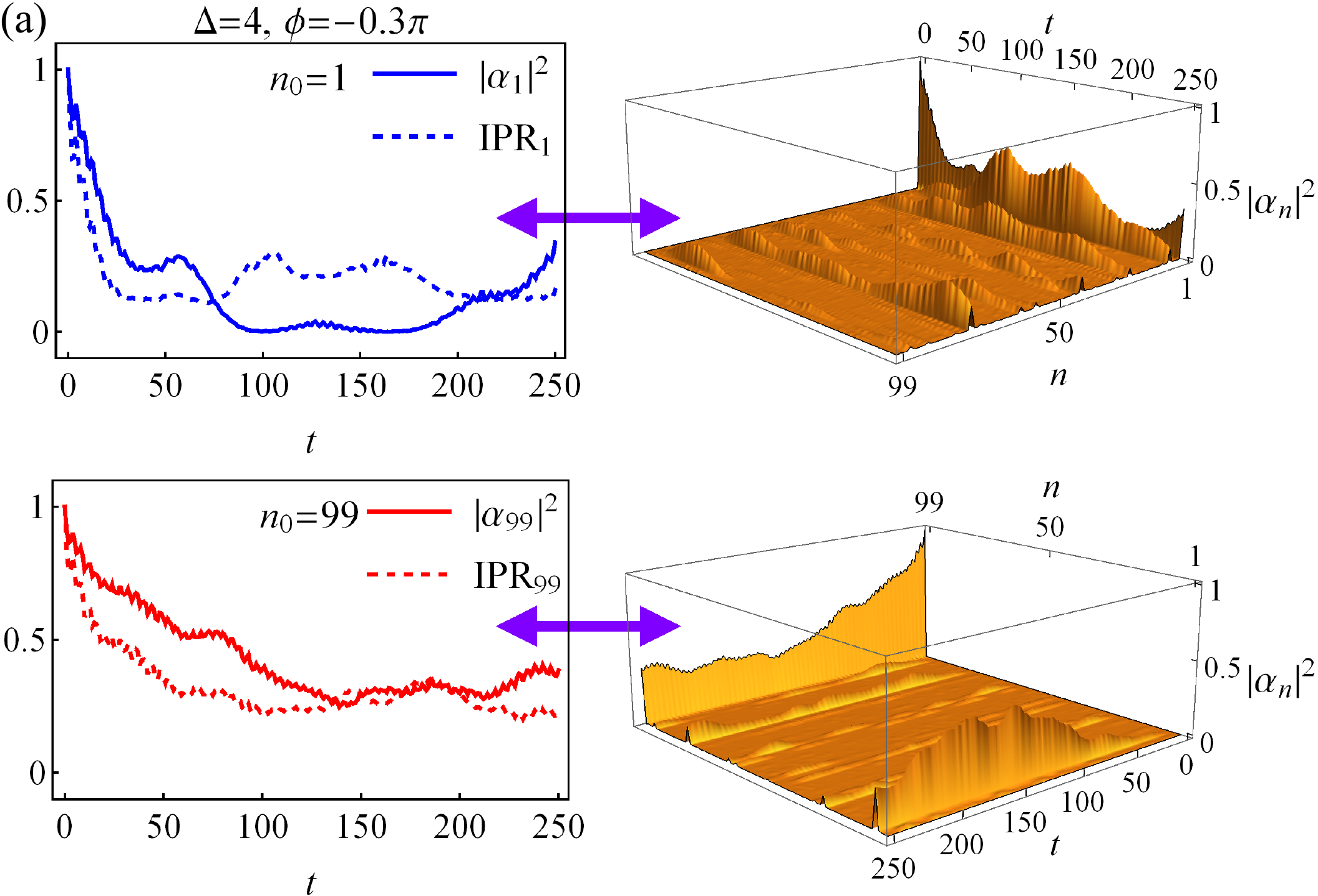}
\includegraphics[width=8.5cm]{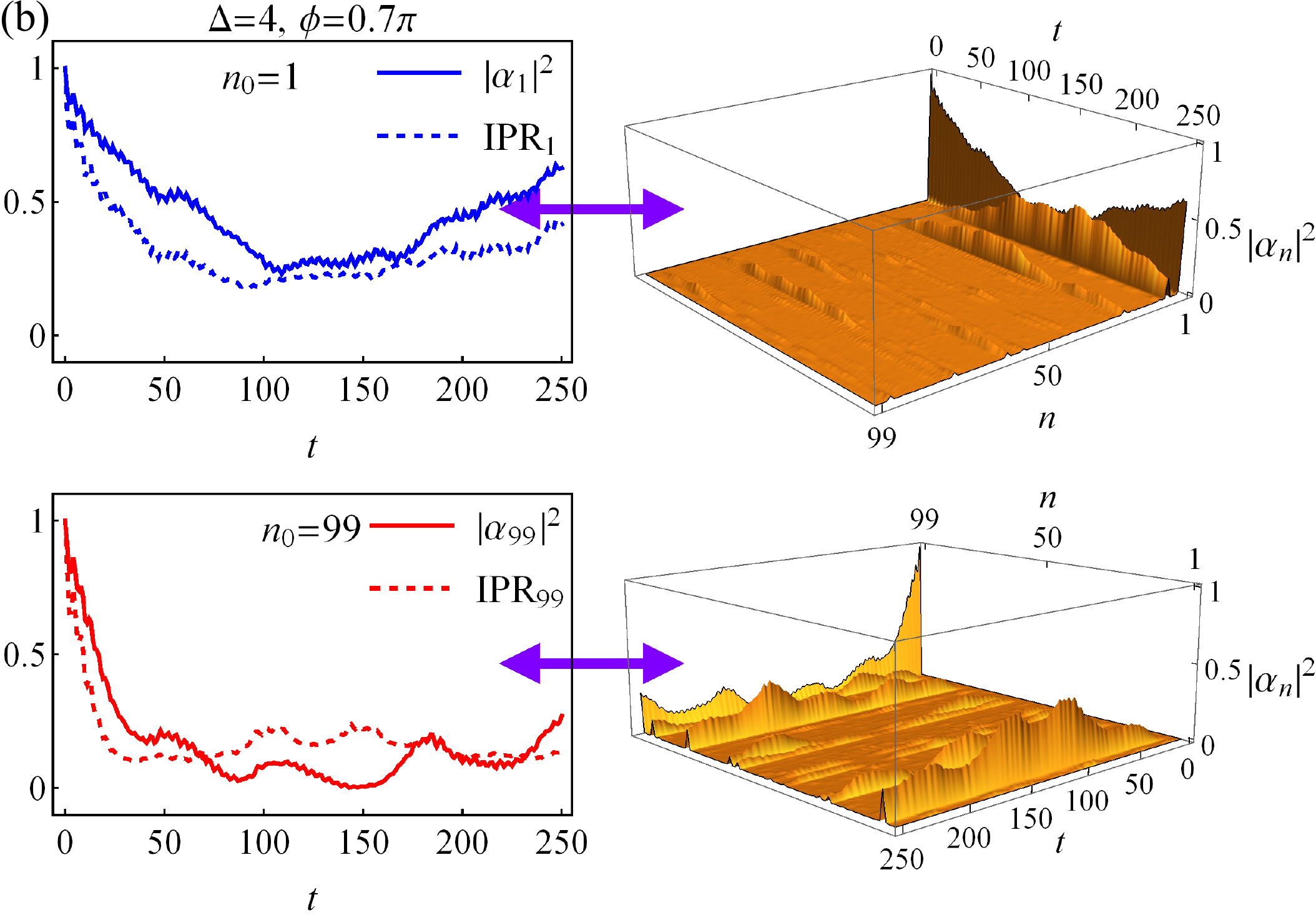}
\caption{(Color online) The evolution of survival probability  for $\phi=-0.3\pi$ (a) and $\phi=0.7\pi$ (b) when $\Delta=4$, as well as $\text{IPR}_{1(99)}$. The other settings are same to those in Fig.\ref{fig:incom-evolution}. }
\label{fig:incom-hopping}
\end{figure}

\begin{figure}
\center
\includegraphics[width=8.5cm]{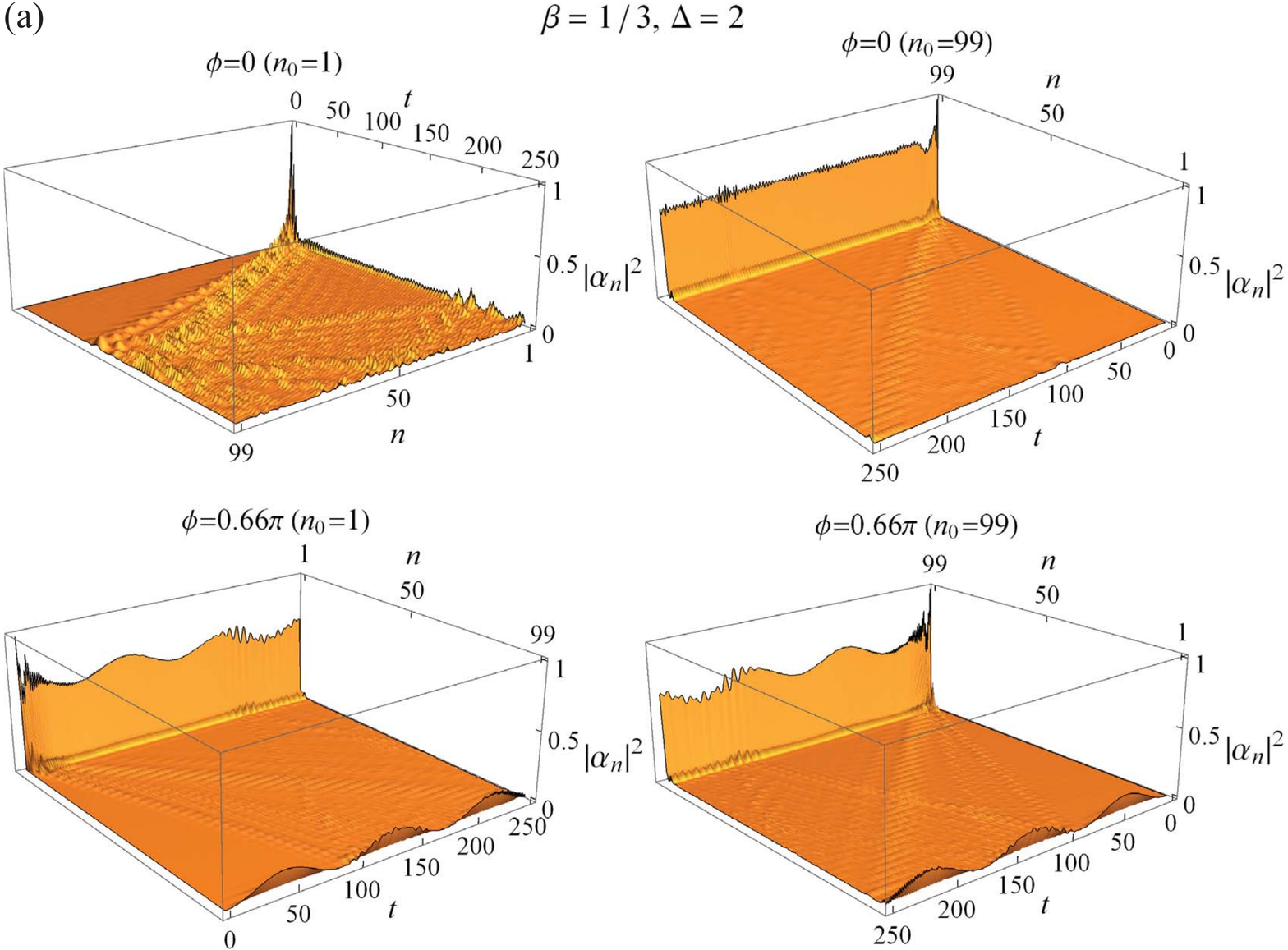}
\includegraphics[width=8.5cm]{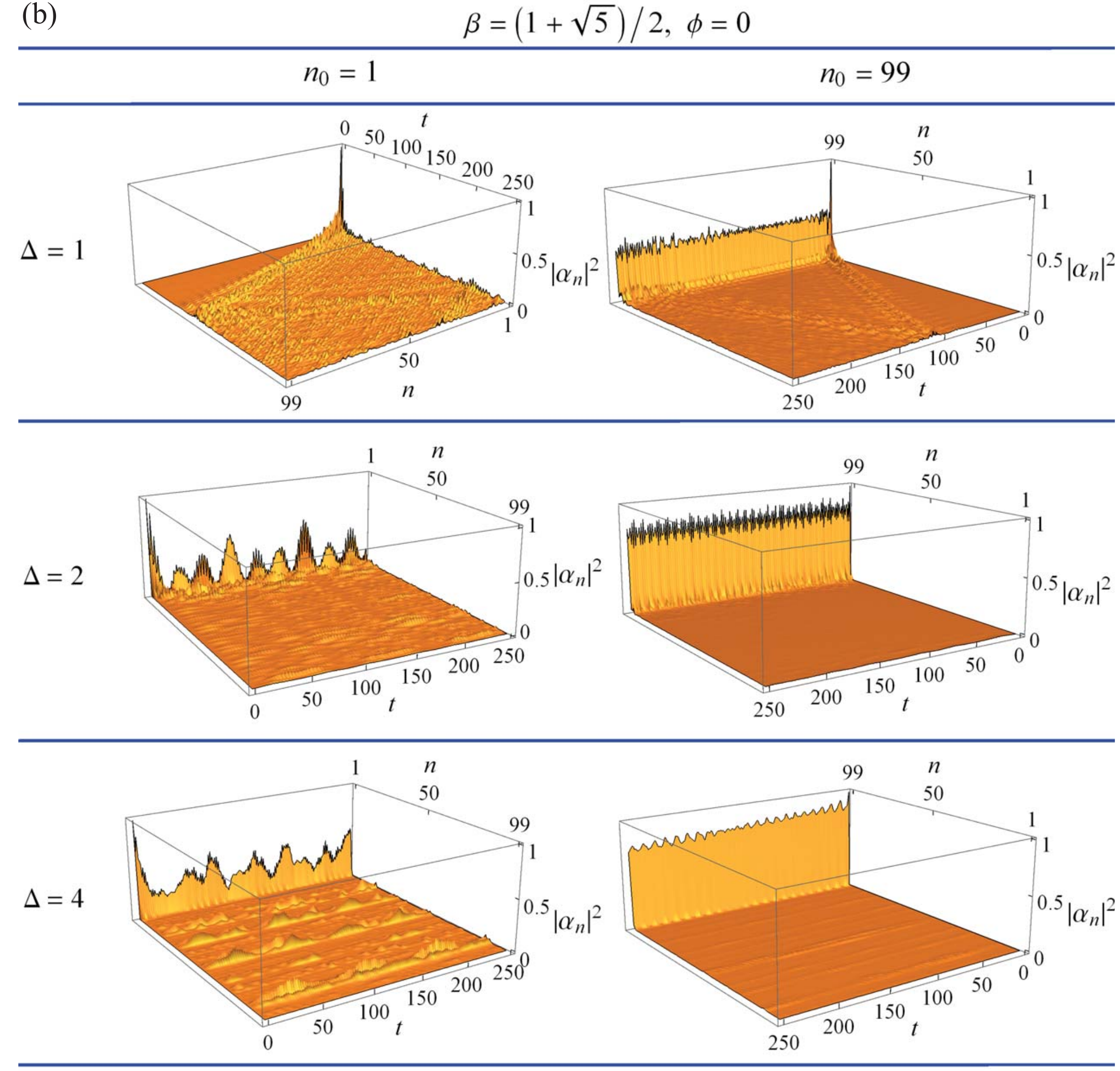}
\caption{(Color online) The plots for the distribution  $\left|\alpha_{n}\right|^2 $ for $\beta=1/3$ (a) and $\beta=\left(1+\sqrt{5}\right)/2$ (b) versus $t$. The selected $\phi$ and $\Delta$ are presented in the plot labels. The other settings are same to those in Fig.\ref{fig:com-evolution} for all plots. }
\label{fig:incom-hopping2}
\end{figure}

\section{The  long-range hopping of excitation}

In order to demonstrate the effect of effective long-range correlation and quasi-disorder, the cases $\phi=-0.3\pi$ and $0.7\pi$ are inspected for $\Delta=4$. There is no DBS or BIC under these circumstances  as shown in Fig.\ref{fig:incom-be}. The survival probability and the corresponding distribution of excitation in system are plotted in Fig.\ref{fig:incom-hopping}. It is clear that the occupation probabilities  of the excitation located on some sites becomes pronounced, except for the initial one. Meanwhile the evolution of IPR also becomes complex. This phenomenon is a result of the interplay of the quasi-disorder and the effective  long-range correlation: The  long-range correlation is devoted to the hopping and dissipation of excitation. Whereas,  the quasi-disorder tends to trap and preserve the excitation against dissipation. Consequently at some moment the excitation is kept  as some site with a significant probability, where the on-site potential is stronger.

However we find that the hopping could be restrained greatly when DBS or BIC  appears. As an exemplification, we examine the  case of $\phi=0$ when $\beta=\left(1+\sqrt{5}\right)/2$, in which there is a BIC with localization at site $n=99$. It is found  for $n_0=1$ that the distribution $\left|\alpha_{n}\right|^2 (n\neq 1)$ becomes pronounced at some sites with the increment of $\Delta$, as shown in Fig. \ref{fig:incom-hopping2} (b). However for  $n_0=99$,  it is clear from Fig.\ref{fig:incom-hopping2}(b) that $\left|\alpha_{n}\right|^2 (n\neq 99)$ tends to disappear even for $\Delta=4$. The phenomenon originates from the strong localization of DBS or BIC, which is protected by the nontrivial topology in $H_S$.

This picture can  also be noted in commensurate case.  As shown in Fig.  \ref{fig:incom-hopping2} (a) for $\phi=0.66\pi$ when $\beta=1/3$, a hopping of excitation can be found \emph{only} between sites $n=1$  and $n=99$.  In contrast, it is absent when there is only one DBS, as shown for $\phi=0$ in Fig.  \ref{fig:incom-hopping2} (a).

%In conclusion the interplay of effective long-range correlation and localization induced by quasi-disorder or edge mode, is responsible for the long-rang hopping of excitation in system.  However when DBS or BIC is occurring in the incommensurate case, the hopping could be compressed greatly.

\section{The Long-time behavior}

\begin{figure}
\center
\includegraphics[width=6cm]{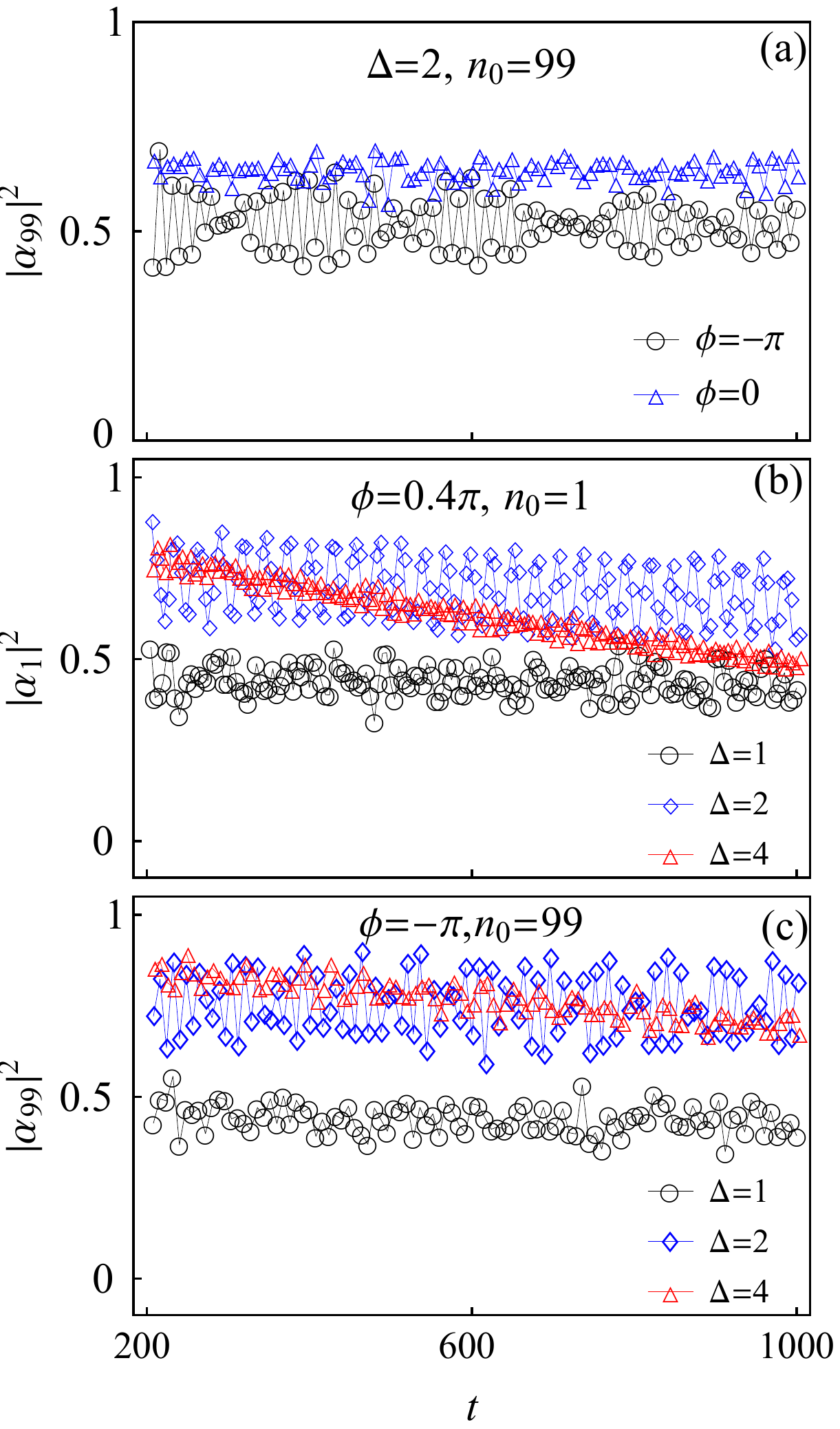}
\caption{(Color online) The  long-time feature of survival probability for excitation with $\beta=1/3$ (a) and $\beta=\left(1+\sqrt{5}\right)/2$ (b, c). The chosen initial states, as well as  $\phi$ and $\Delta$, are presented in the plot labels.  $N=99$,  $s=1$, $\eta=0.1$ and $\omega_c=10$ are chosen for all plots. }
\label{fig:longtime}
\end{figure}

Although we claim  that DBS or BIC could determine the steady behavior of system, an exception can be found. As for $\phi=0.4\pi$ and $-\pi$ with  $\beta=\left(1+\sqrt{5}\right)/2$,  the survival probability of excitation declines very slowly when  $\Delta=4$, even if the initial state is overlapped  with DBS or BIC, as shown in Fig. \ref{fig:longtime} (b) and (c). We find that this declination cannot be attributed to computational errors. In contrast it does not occur for $\beta=1/3$, as shown in Fig. \ref{fig:longtime} (a), as well for $\Delta=1$ when $\beta=\left(1+\sqrt{5}\right)/2$  shown in Fig. \ref{fig:longtime} (b) and (c). For these two cases, the system is extendible or in delocalized phase. For longer time evolution, the numerical evaluation becomes exhaustive and thus unreliable because of the accumulation of computational error.

Unfortunately we cannot determine the  reason for this declination because of the difficulty  of deciding all bound states. As for this phenomena is absent when the system is extendible or delocalized, a possible understanding might be the influence of the bound states in band.  These states also become much localized with the increment of disorder in $H_S$. As a consequence they would show  non-negligible contribution to the evolution after a long time.

\section{Conclusion}

In conclusion,  the bound states and  their influence on the population evolution are investigated in a one-dimensional tight-binding atomic chain. Each site of the chain is  coupled to an  environment and all sites share a common environment. By solving the Schr\"{o}dinger equation in the limit of a single excitation,  three special kinds of bound states are identified. The first is the DBS, which corresponds to a single negative eigen-energy  with  finite gap from the continuum. It is concluded from the calculations that the system on DBS does not decay, and it has  similar localization features to the edge mode of the system. An additional DBS is found in the gap when the system is  commensurate, which is extendible  and can be understood as  the bath-induced transition of the state in band. The situation changes when the system is incommensurate due to the intrinsic localization in system, which  prevents the system  being excited due to its couplings to the environment.

The second is a bound  state in continuum, which is connected intimately to the edge mode with positive energy and also exhibits zero decay rate. The robustness of BIC could be attributed to the nontrivial topology of the system.   The third is a single special bound state of the lowest energy. Different from the first two bound states,  it is extendible and displays a certain probability to decay.  Moreover it depends sharply  on the size of system and the properties of the bath.

The time evolution of a single excitation is simulated in order to explore the influence of the bound states. It is concluded  that the bound states are predominant for the population evolution. When the system is extendible or delocalized, the excitation  becomes stable against dissipation provided  the initial state  overlaps  with DBS or BIC. However if the overlapping is zero,  the evolution is  dissipative  and the information of initial states will be  erased finally. The situation changes  for incommensurate systems with  strong quasi-disorders (for example, $\Delta=4$), the occupation probabilities of the excitation decrease slowly, even if the initial state  overlaps with DBS or BIC. Furthermore a significant recurrence of  survival probabilities for the excitation can be found when the initial state overlaps with neither DBS nor BIC. These two features may be understood as the interplay between  localizations in the system and the effective long-range correlation induced by the  bath. Another  important consequence of this interplay is the long-range hopping of the single excitation of the system, which makes the excitation hop to a different site from an initial one. We note that the hopping can also happen between two localized DBSs  in commensurate cases, as shown for $\phi=0.66\pi$ in Fig. \ref{fig:com-evolution}(a6) and (b6).

An open question is the effect of interactions between atoms on the prediction. It is known that the competition between interactions and disorders is responsible for the  many-body localization transition in AAH model \cite{iyer13}. Recall that the interatomic interaction might destroy the localization, the edge mode in the system could be changed. Moreover, the  bound states in open systems amount to  an effective trap potential \cite{shi16, shi18}, which prevents excitation from decaying. Hence  when the interatomic interactions are involved,  the competition between the effective trapping  and  interatomic interactions  would intrigue interesting feature. When the trapping is predominant, the excitation could be preserved  in the system. Otherwise, the excitation dissipates. Due to the complicated and involved  calculation for multi-excitation bound states \cite{shi16}, we left the related  discussion  in the future work.

\section*{Acknowledgments}
HZS acknowledges the support of National Natural Science Foundation of China (NSFC) under Grant No.11705025. MQ acknowledges the support of National Natural Science Foundation of China (NSFC) under Grant No. 11805092. HTC and XXY acknowledges the financial support of NSFC under Grants No. 11775048, No. 11534002 and No. 11947405.

\newpage
\widetext
\renewcommand\thefigure{A\arabic{figure}}
\renewcommand\theequation{A\arabic{equation}}
\setcounter{equation}{0}
\setcounter{figure}{0}

\section*{Appendix A}

In this appendix, the exact solutions to  Eq. (\ref{bs}) is presented  for $\beta=p/q$ under periodic boundary condition. Assume $N=Lq$, and then $H_S$ can be written as
\be
H_S&= &\sum_{x=1}^L \left(c_1^{\dagger}, c_2^{\dagger},\cdots, c_q^{\dagger}\right)_x \left(\begin{array}{cccc} \Delta\cos\left(\frac{2\pi p}{q} +\phi\right) & J & 0 & \cdots \\
J & \Delta\cos\left(\frac{4\pi p}{q} +\phi\right) & J & 0 \\
\vdots & \vdots & \ddots & \vdots \\
0 & \cdots & J & \Delta\cos\left(\phi\right) \end{array}\right)\left(\begin{array}{c} c_1 \\ c_2\\ \vdots \\ c_q\end{array}\right)_x +\nonumber\\
&&\sum_{x=1}^L \left(c_1^{\dagger}, c_2^{\dagger},\cdots, c_q^{\dagger}\right)_x \left(\begin{array}{cccc}
0 & 0& \cdots & 0  \\ \vdots & \ddots & \vdots & \vdots\\
 1 & 0 &\cdots &  0
\end{array}\right)\left(\begin{array}{c} c_1 \\ c_2\\ \vdots \\ c_q\end{array}\right)_{x+1} + h.c.
\ee
By Fourier transformation $c_x= \frac{1}{\sqrt{L}} \sum_{\lambda=1}^L a_{\lambda} e^{i2\pi \lambda x/L}$, then
\be
H_S= \sum_{\lambda=1}^L \left(a_1^{\dagger}, a_2^{\dagger},\cdots, a_q^{\dagger}\right)_{\lambda} \left(\begin{array}{ccccc} \Delta\cos\left(\frac{2\pi p}{q} +\phi\right) & J & 0 & \cdots & e^{i2\pi q \lambda/L}\\
J & \Delta\cos\left(\frac{4\pi p}{q} +\phi\right) & J & 0& \cdots \\
\vdots & \vdots & \ddots & \vdots & \vdots \\
e^{-i2\pi q \lambda /L}& 0 & \cdots & J & \Delta\cos\left(\phi\right) \end{array}\right)\left(\begin{array}{c} a_1 \\ a_2\\ \vdots \\ a_q\end{array}\right)_{\lambda}.\nonumber
\ee
As for $H_{int}$,
\be
H_{int}&=&\sum_{k, x}g_k b_k \left(c_1^{\dagger}, c_2^{\dagger},\cdots, c_q^{\dagger}\right)_x+ g_k^* b_k^{\dagger} \left(\begin{array}{c} c_1 \\ c_2\\ \vdots \\ c_q\end{array}\right)_x \nonumber\\
&&\Rightarrow  \frac{1}{\sqrt{L}}\sum_{k}g_k b_k \left(a_1^{\dagger}, a_2^{\dagger},\cdots, a_q^{\dagger}\right)_{\lambda=0}+ g_k^* b_k^{\dagger} \left(\begin{array}{c} a_1 \\ a_2\\ \vdots \\ a_q\end{array}\right)_{\lambda=0}.
\ee

\begin{figure}
\center
\includegraphics[width=12cm]{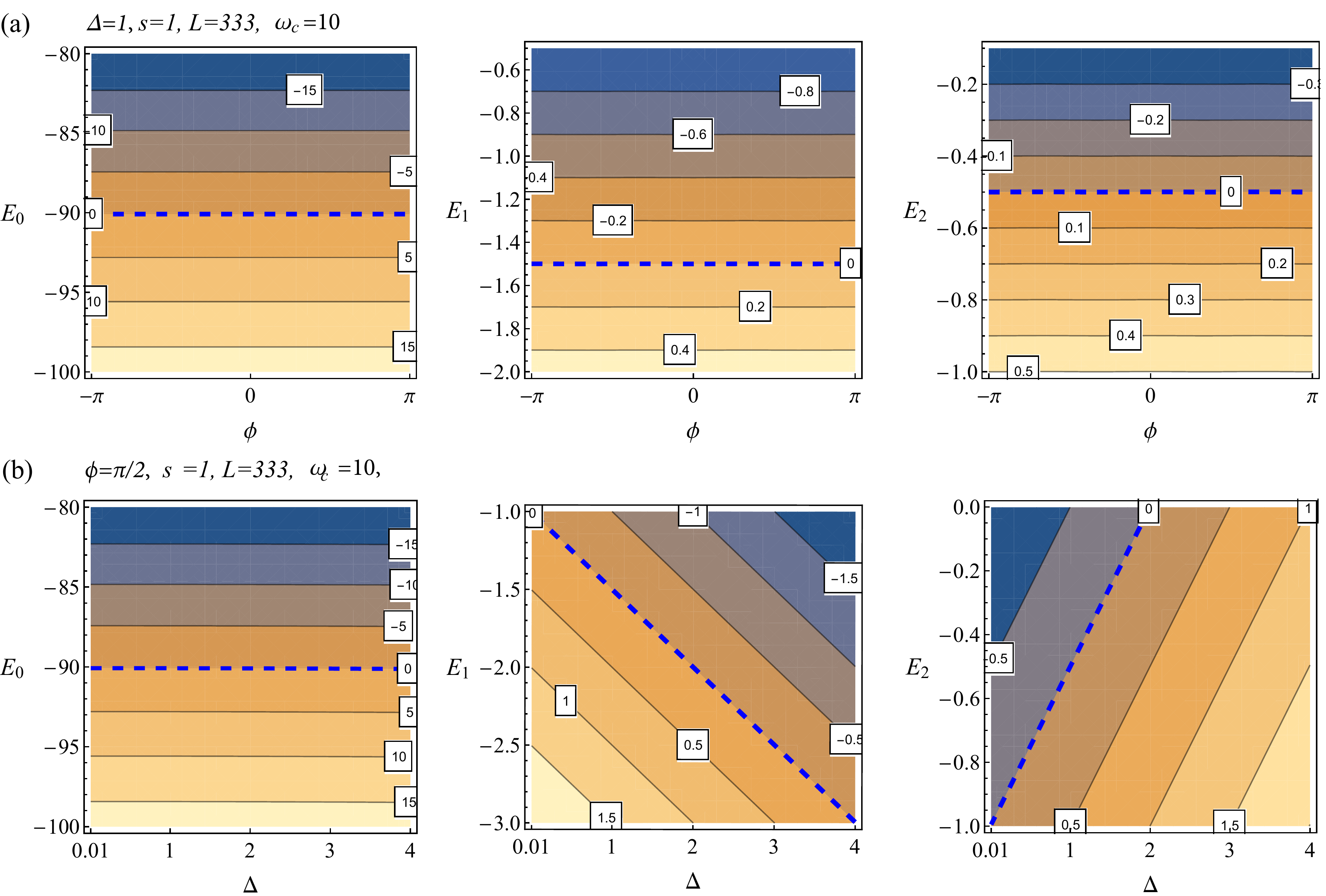}
\includegraphics[width=12cm]{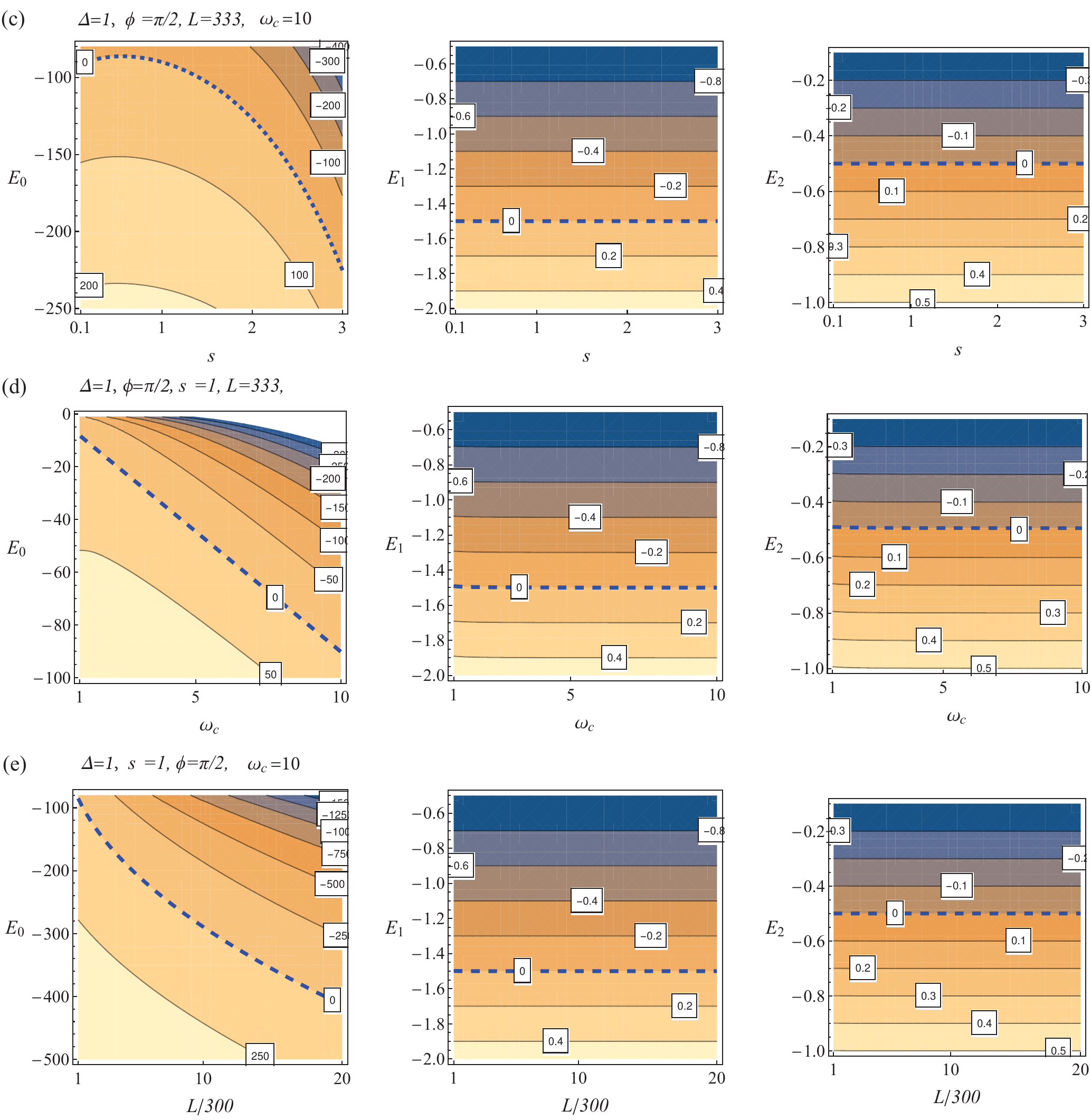}
\caption{(Color online) The plots of the numerical solutions to Eq.  (\ref{solution}) versus the different parameters in system and bath. $\eta=0.1$ is chosen for all plots.   }
\label{fig:appendix}
\end{figure}

Considering  a single excitation for $\lambda=0$, the eigenfunction  $\ket{\psi}_E$ can be written as
\be
\ket{\psi}_E= \left(\sum_{n=1}^q \alpha_n a^{\dagger}_n \ket{0}_n \right)\otimes \ket{0}^{\otimes M} +\ket{0}^{\otimes q} \otimes \left(\sum_{k=1}^{M} \beta_k b_k\ket{0}_k\ket{0}^{\otimes (M-1)} \right).
\ee
As for  $p=1$ and $ q=3$, substitute $\ket{\psi}_E$ into Eq.  (\ref{bs}) and eliminate the degree of freedom of bath. One can obtain the equation
\be
E^3- 3 d(E) E^2 - \left(3 + 6 d(E) + \frac{3}{4}\Delta^2\right)E - \left(2 + 3d(E) - \frac{3a}{4}\Delta^2 + \frac{\Delta^3}{4}\cos3\phi\right)=0,\nonumber
\ee
where $d(E)=\frac{1}{L}  \int_0^{\infty} \frac{J(\omega)}{E-\omega} \text{d}\omega $. By solving the above equation, three relations can be found
\be\label{solution}
E_0&=& \sqrt{4\left[1 +d(E_0)\right]^2 + \Delta^2} \cos\left(\theta_{\phi}+\frac{2\pi}{3}\right) +d(E_0);\nonumber\\
E_1&=& \sqrt{4\left[1 +d(E_1)\right]^2 + \Delta^2} \cos\left(\theta_{\phi}+\frac{4\pi}{3}\right) +d(E_1);\nonumber\\
E_2&=& \sqrt{4\left[1 +d(E_2)\right]^2 + \Delta^2} \cos\theta_{\phi} +d(E_2),
\ee
where
\be
\theta_{\phi}= \frac{1}{3}\arccos \left\{\frac{\left[1 +d(E)\right]^3 + \frac{1}{8}\Delta^3\cos3\phi}{\sqrt[3]{\left[1 +d(E)\right]^2 + \Delta^2/4}}\right\}.
\ee
$E_0, E_1, E_2$ correspond to three real solutions, which are plotted for different parameters by  blue dashed lines in Fig.\ref{fig:appendix}. We find that $E_0$ shows significant dependence on the properties of the bath and  the system size $L$, and thus it is extensive. In contrast, both $E_1$ and $E_2$ are determined completely by the properties of system, and thus are intensive.

\begin{figure}[t]
\center
\includegraphics[width=15cm]{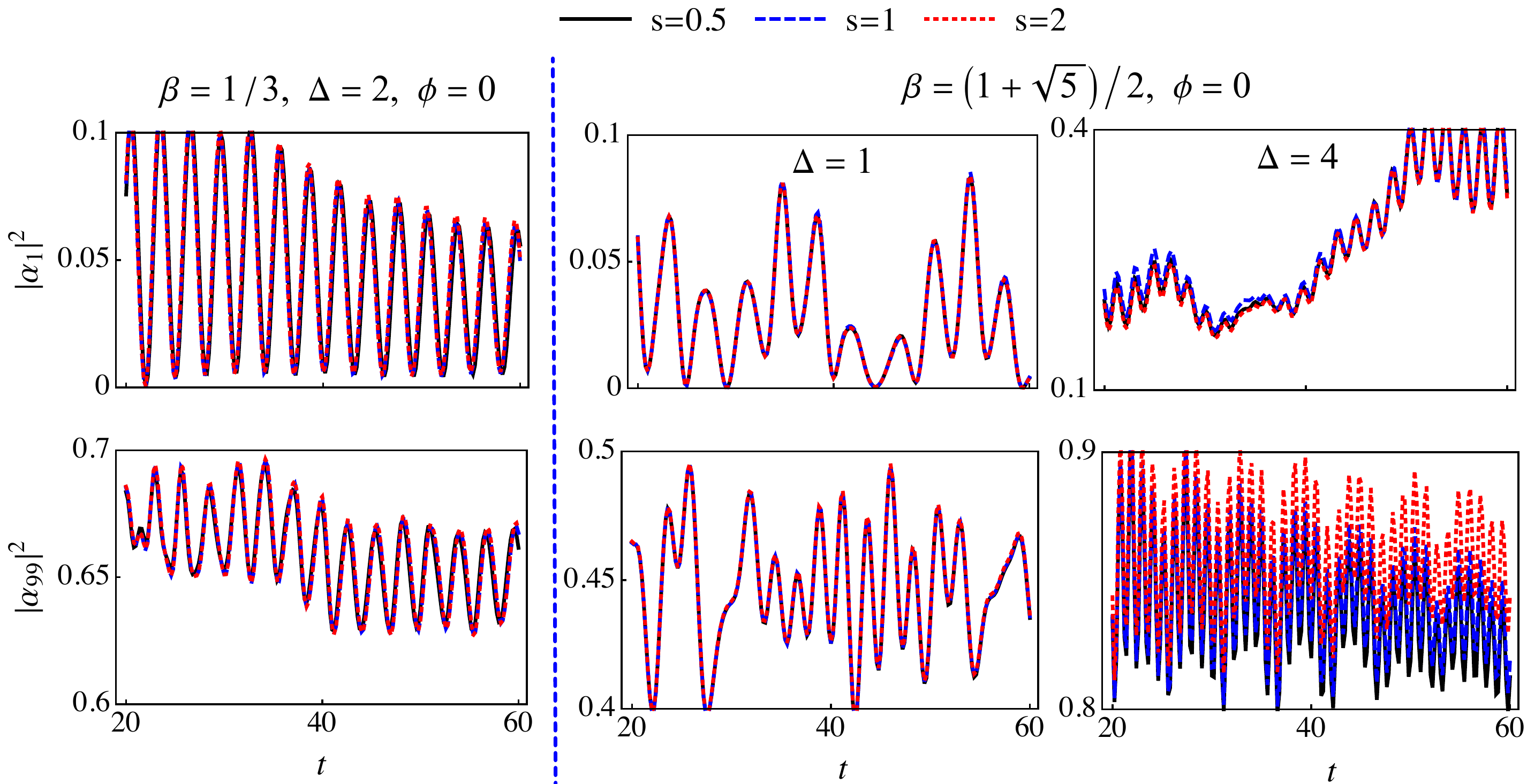}
\caption{(Color online) The plots of survival probability $\left|\alpha_{1(99)}\right|^2$  for different $s$, when the excitation is initially at $n_0=1$ and $n_{0}=99$ respectively. $N=99$ $\eta=0.1, \omega_c=10$ for all plots. }
\label{fig:s}
\end{figure}

Actually the three levels  $E_0$ and $E_1, E_2$ characterize the main feature of the bound state in main text. $E_0$ corresponds to the minimal  solution to Eq. \eqref{boundeqn}, which is extended and has a finite probability of spontaneous emission.  However $E_1$ and $E_2$  have correspondence to DBS.  In Fig.\ref{fig:s}, the evolution of excitation initially at $n_0=1, 99$ are plotted  for different $s$. It is apparent that the survival probability is insensitive to the value of  $s$.

\renewcommand\thefigure{B\arabic{figure}}
\renewcommand\theequation{B\arabic{equation}}
\setcounter{equation}{0}
\setcounter{figure}{0}

%%%%%%%%%%%%%%%%%%%%%%%%%%%%%%%%

\section*{Appendix B}

The inverse participation ratio (IPR) is a general measure of the localization of state. For  state $\ket{\psi}=\sum_{n=1}^N \alpha_n \ket{n}$, where $\ket{n}$ denotes the occupation of the $n$-th site, and $N$ is the number of site,  IPR is defined  as
\be
\text{IPR}_{\psi}=\sum_{n=1}^N \left|\alpha_n\right|^4.
\ee
IPR has the minimum $1/N$ only if $\left|\alpha_n\right|^2=1/N$ for any $n$, which means that the distribution of excitation is uniform, and thus $\ket{\psi}$ is extended. While IPR has the maximum 1 only if $\left|\alpha_n\right|^2=1$ for a special $n$, which means that excitation  can appear only at site $n$, and thus  $\ket{\psi}$ is localized completely.

\begin{figure}
\center
\includegraphics[width=5.5cm]{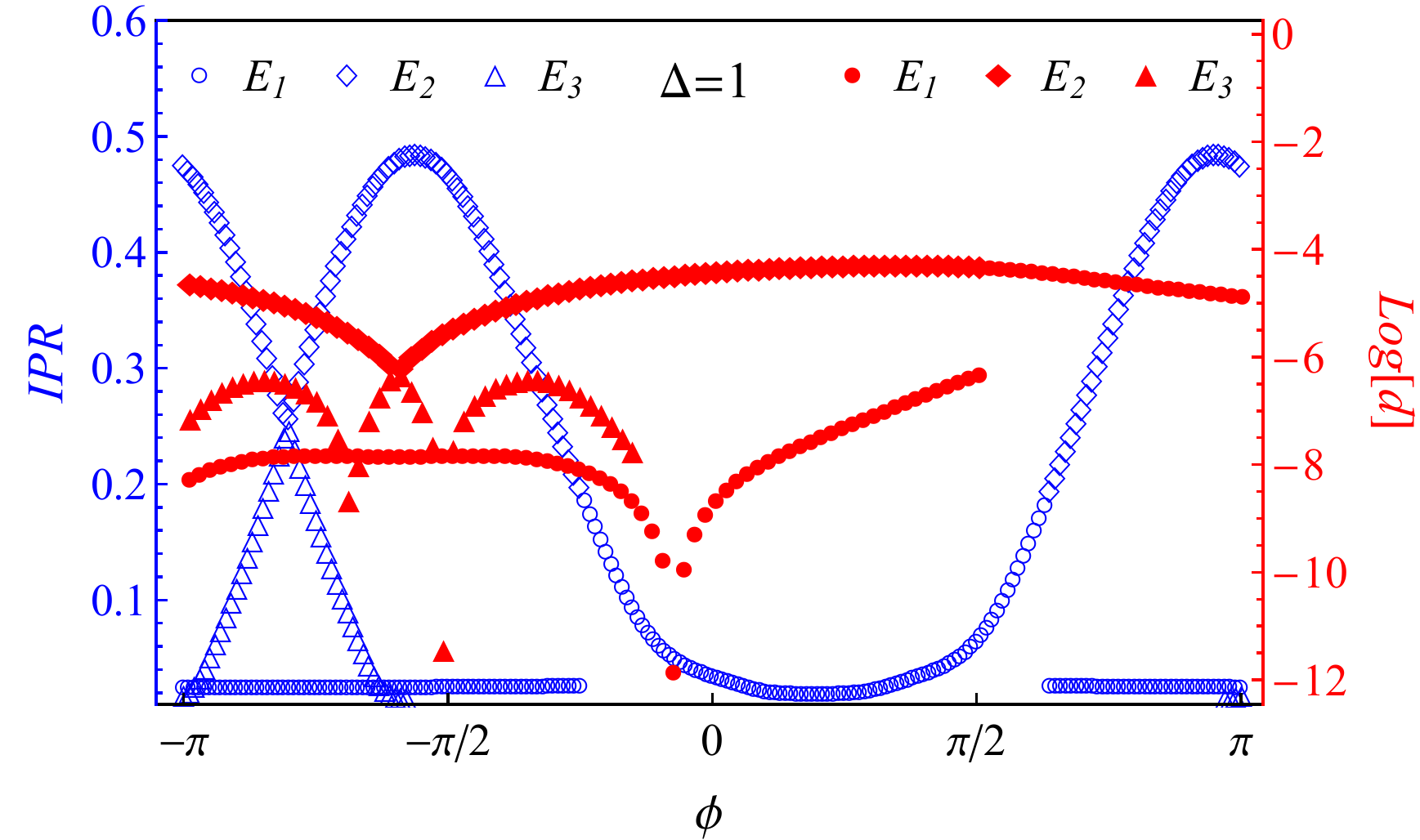}
\includegraphics[width=5.5cm]{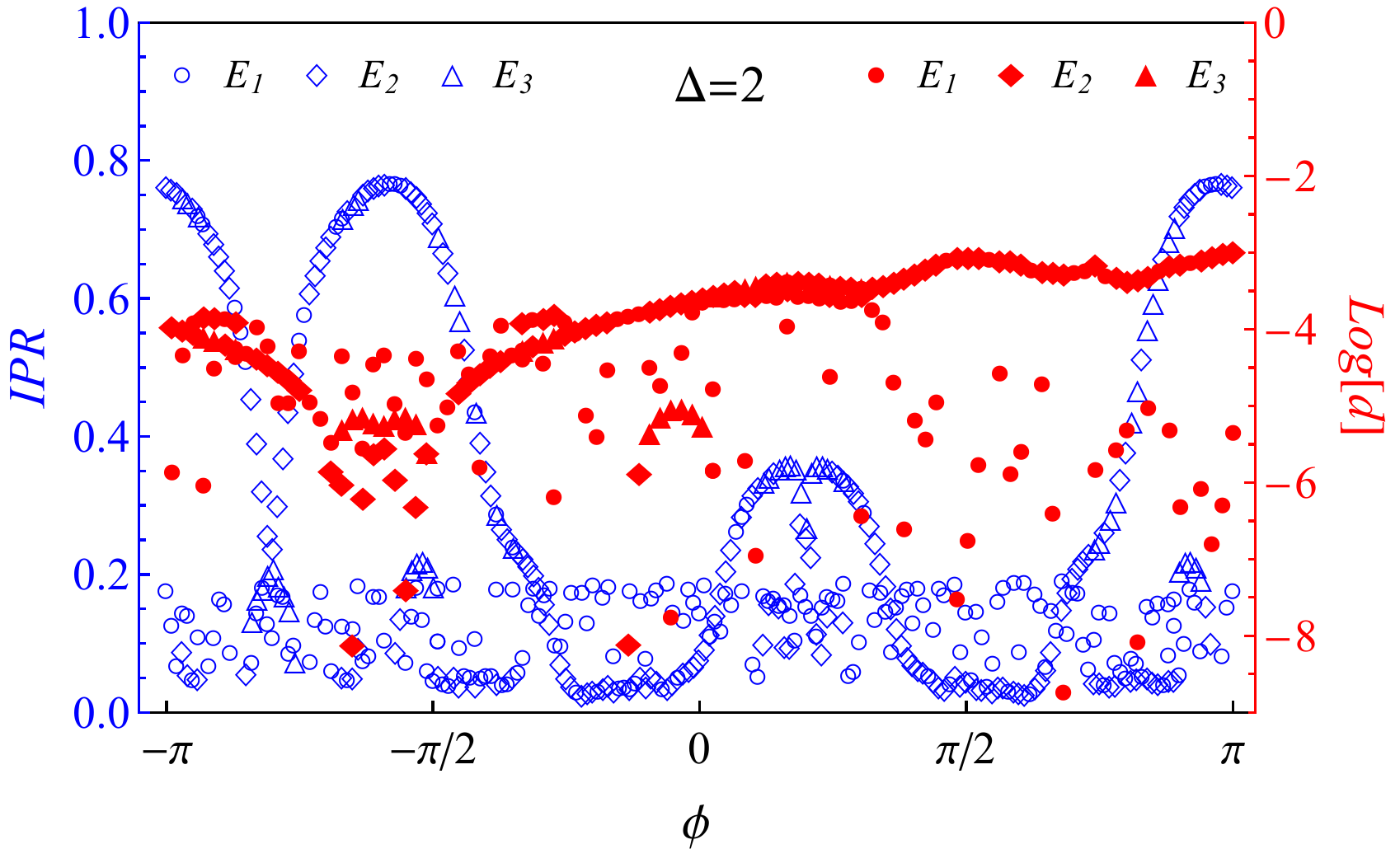}
\includegraphics[width=5.5cm]{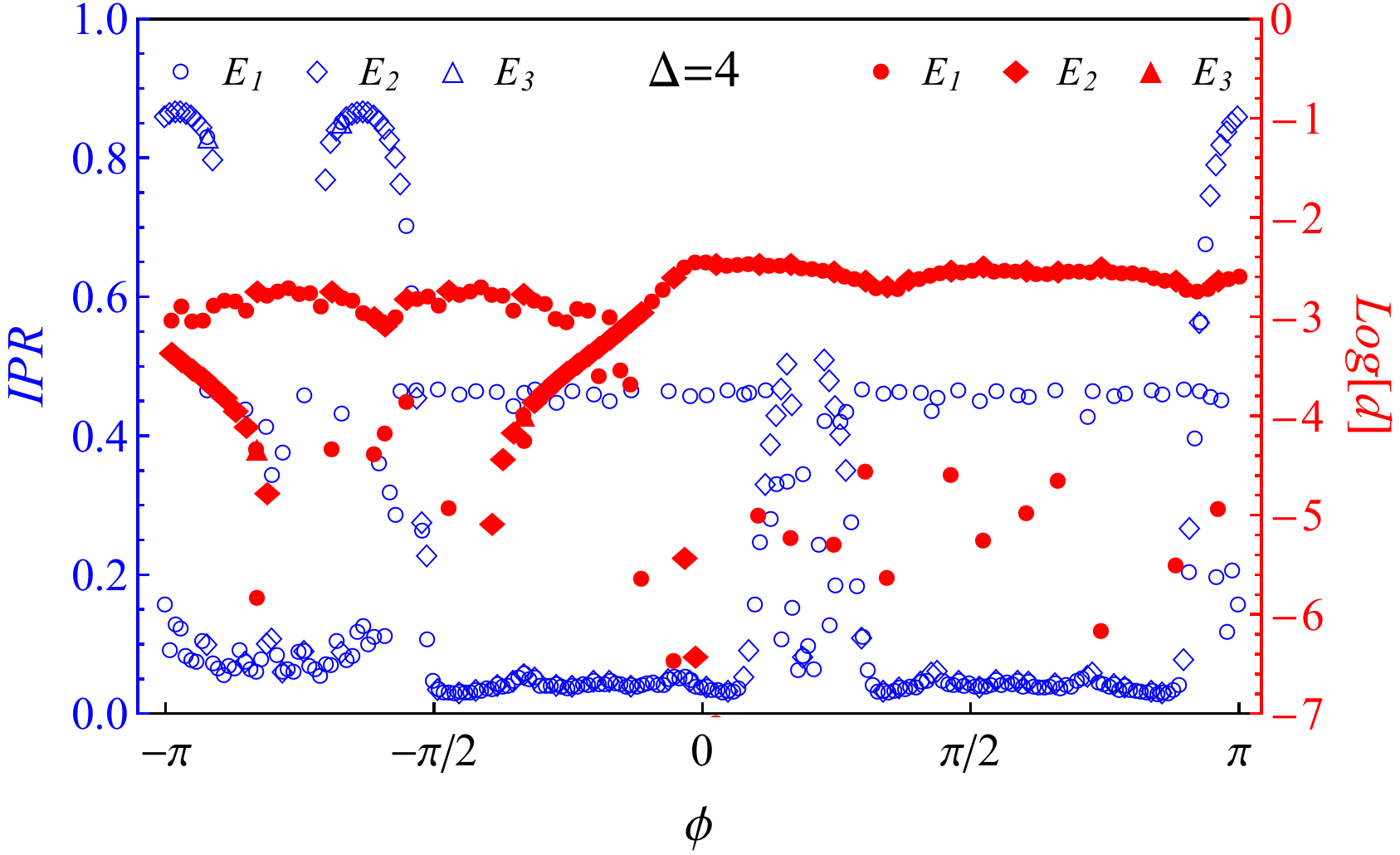}
\caption{(Color online) Plots of IPR (blue empty symbols) and  $d$ (red-solid symbols) for DBS when $\beta=\left(1+\sqrt{5}\right)/2$ and $\Delta=1, 2, 4$, respectively.  The parameters are chosen as the same in Fig. \ref{fig:incom-be}. The labels of $E_1$, $E_2$ and $E_3$ denote the levels of DBS, plotted in Fig. \ref{fig:incom-be}, in increscent order. }
\label{fig:incom-bs}
\end{figure}

In Fig. \ref{fig:incom-bs}, IPR and corresponding $d$ are plotted for different $\Delta$s when $\beta=\left(1+\sqrt{5}\right)/2$. It is clearly concluded that the DBS is localized. Moreover IPR is enhanced with the increment of $\Delta$, which means that the system becomes more localized.

\begin{figure}[t]
\center
\includegraphics[width=15cm]{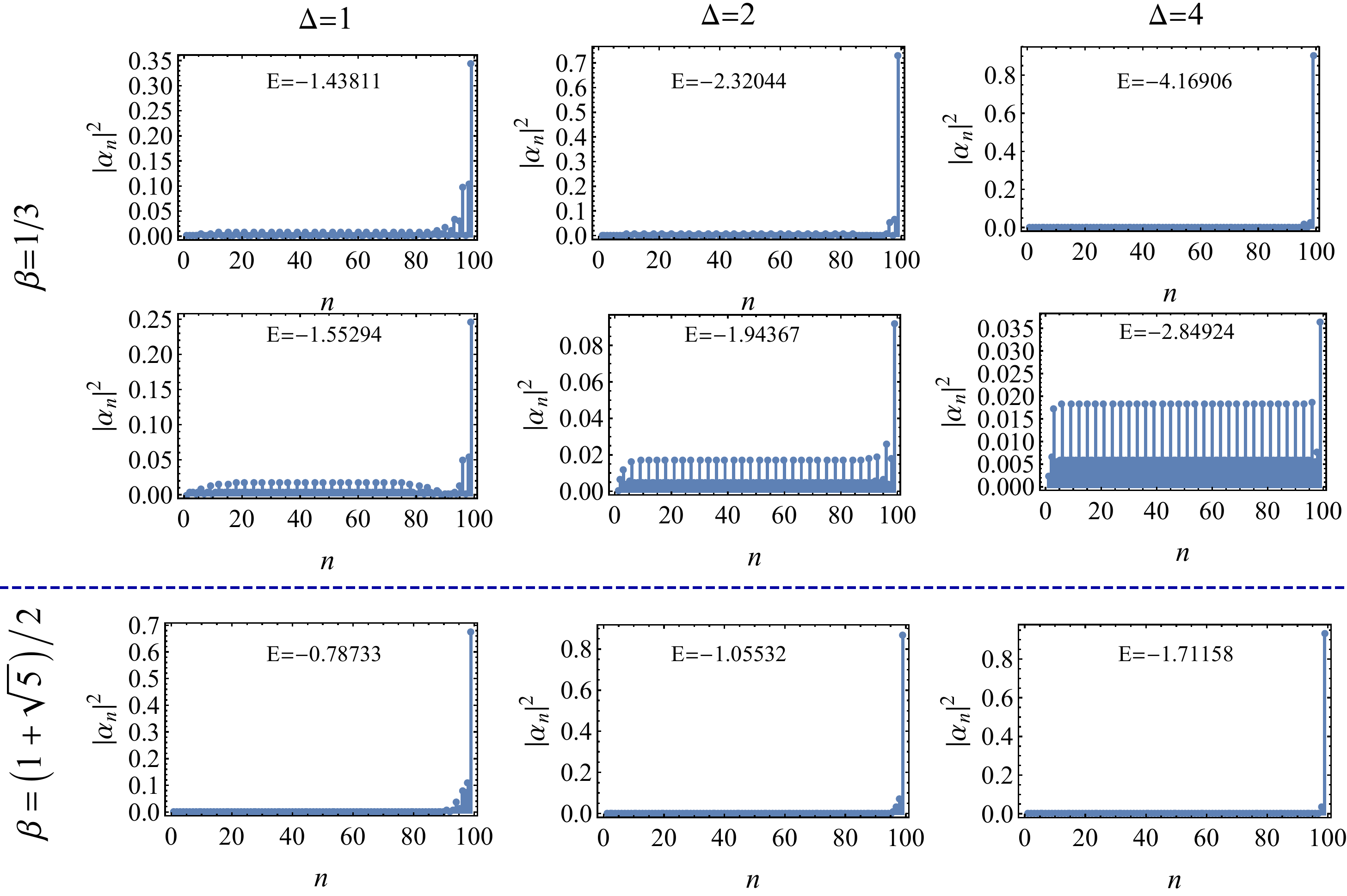}
\caption{(Color online) The site-distribution  $\left|\alpha_n\right|^2$ for DBS when $\phi=-\pi$ versus $\Delta=1,2,4$.  $N=99$, $s=1, \eta=0.1, \omega_c=10$ are chosen for all plots.}
\label{fig:distribution}
\end{figure}

We note that the IPR of DBS is always smaller than 1. The reason is the interplay between the localization, which tends to localize  the excitation in system, and the effective long-range correlation in atomic sites,  which tends to delocalize  the excitation instead. In Fig. \ref{fig:distribution}, the distribution $\left|\alpha_n\right|^2$ of excitation in DBS  is shown for different $\Delta$ when $\phi=-\pi$ as an exemplification.  When $\beta=1/3$, there is two DBSs. One  corresponds to the renormalized edge state, and thus show strong localization. The other comes from the transition of state in band, and thus is extended.  It is clear for the former that  the distribution  becomes much pronounced at end site $n=99$  with the increment of $\Delta$,  as shown in the upper row  in  Fig. \ref{fig:distribution}.  However for the latter  it tends to be multipeaked with the increment of $\Delta$, as shown in the middle row in Fig. \ref{fig:distribution}. As for  $\beta=\left(1+\sqrt{5}\right)/2$,  the value of $\Delta$ characterizes the strength of disorder in  system. Thus it is not surprising that the localization of DBS is enhanced  with the increment of $\Delta$, as shown by the bottom row in Fig.\ref{fig:distribution}.

%The reason is that  the onsite potential is inclined to keep the original property of state invariant. Note that the emergent bound state is excited from the band and thus is extended in nature, it is not strange for the enhancement of extensity in the emergent bound state.

\renewcommand\thefigure{C\arabic{figure}}
\renewcommand\theequation{C\arabic{equation}}
\setcounter{equation}{0}
\setcounter{figure}{0}

\section*{Appendix C}

In this appendix,  we demonstrate the existence of BIC analytically. For this purpose, we first  diagonalize the system Hamiltonian  as $H_S=\sum_{i=1}^{N} \epsilon_i \eta^{\dagger}_i\eta_i$, where $\eta_i=\sum_n \gamma_{in}^{*}c_n$. The array $\left(\gamma_{i1}, \gamma_{i2},\cdots, \gamma_{iN}\right)^{T}$ denotes the $i$-th eigenstate of Eq. (\ref{hs}). Then the total Hamiltonian can be rewritten as
\be
H=\sum_{i=1}^{N} \epsilon_i \eta^{\dagger}_i\eta_i + \sum_k \omega_k b_k^{\dagger}b_k + \sum_{i,k}g^*_{ik} \eta_i b_k^{\dagger} + g_{ik} \eta^{\dagger}_i b_k.
\ee
where $ g_{ik}=g_k\sum_n\gamma_{in}$. For a arbitrary state $\ket{\psi(t)}=\left(\sum_i \alpha_i(t)\eta_i^{\dagger}\ket{0}_i\right)\ket{0}^{\otimes M} + \ket{0}^{\otimes N}\left(\sum_k \beta_k(t)b_k^{\dagger}\ket{0}_k\right)$, the evolution equation can be written as
\be
\mathbbm{i}\frac{\partial \alpha_i(t)}{\partial t}&=&\alpha_i(t)\epsilon_i -\mathbbm{i}\left(\sum_n\gamma^*_{in}\right) \sum_j \left(\sum_n\gamma_{jn}\right)\int_0^{\tau}\text{d}\tau \alpha_j(t)\sum_k\left|g_k\right|^2e^{- \mathbbm{i}\omega_k t}\nonumber\\
&=&\alpha_i(t)\epsilon_i -\mathbbm{i}\left(\sum_n\gamma^*_{in}\right) \sum_j \left(\sum_n\gamma_{jn}\right)\int_0^{\tau}\text{d}\tau \alpha_j(t)\int_0^{\infty}J(\omega)e^{- \mathbbm{i}\omega t},\nonumber
\ee
where we have assumed that the excitation is located initially in  system, and thus $\beta_k(0)=0$. By Laplace transformation $G_i(z)=\int_0^{\infty}\text{d}t \alpha_i(t) e^{-zt}$, the equation above can be rewritten as
\be\label{appendixB}
\left(\mathbbm{i}z - \epsilon_i \right)G_i(z)- \Sigma(z)\left(\sum_n\gamma^*_{in}\right) \sum_j \left(\sum_n\gamma_{jn}\right)G_j(z)= \mathbbm{i} \alpha_i(0),
\ee
where $\Sigma(z)= \int_0^{\infty}\frac{J(\omega)}{\mathbbm{i}z-\omega}$ is the self-energy. Then we obtain a linear system of equations for $G_i(z)$, for which the solution can be expressed as
\be
G_i(z)= \frac{Det(B_i)}{Det(A)}.
\ee
The element of coefficients matrix $A$ is $A_{ij}= \left(\mathbbm{i}z - \epsilon \right)\delta_{ij}- \Sigma(z)\left(\sum_n\gamma^*_{in}\right)\left(\sum_n\gamma_{jn}\right)$, $B_i$ denotes the modified $A$ with the $i$-th column replaced by $\left(\alpha_1(0),\alpha_2(0), \cdots, \alpha_N(0)\right)^T$.

\begin{figure}
\center
\includegraphics[width=5.5cm]{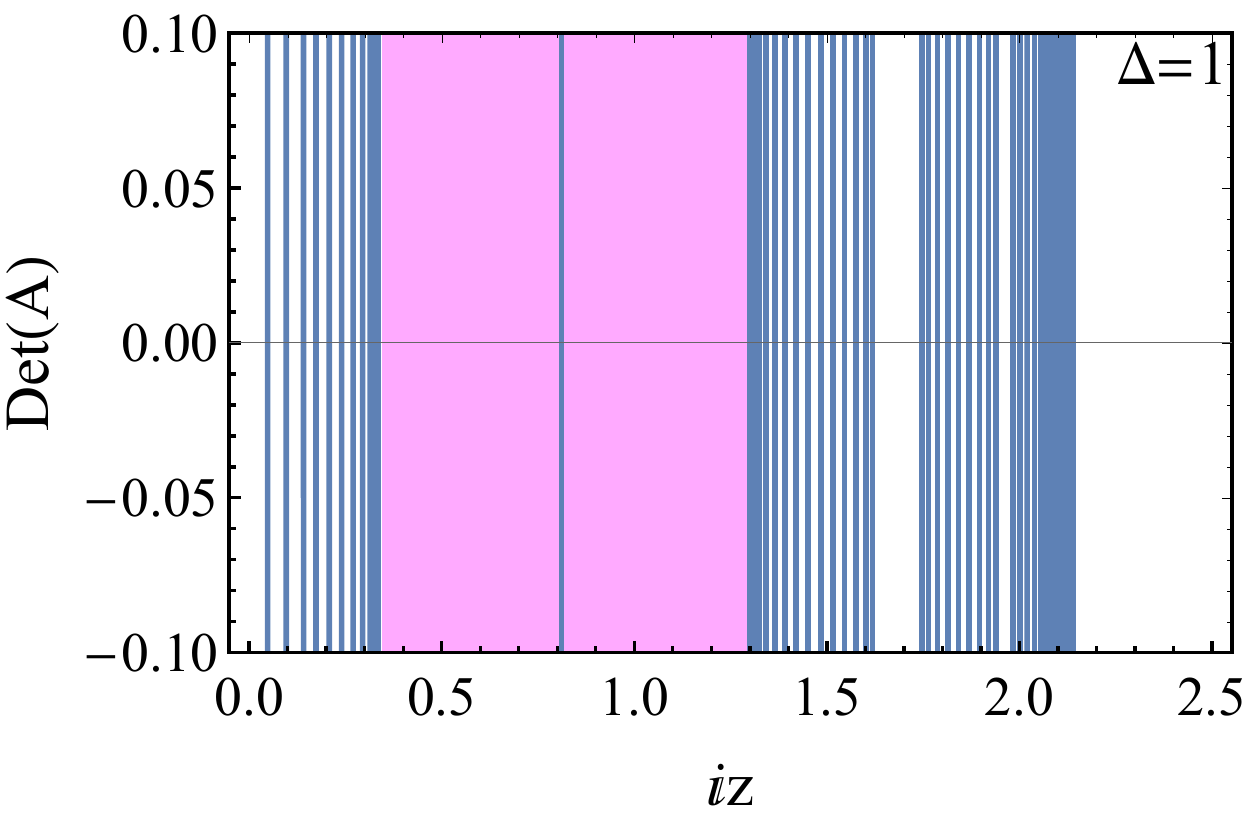}
\includegraphics[width=5.5cm]{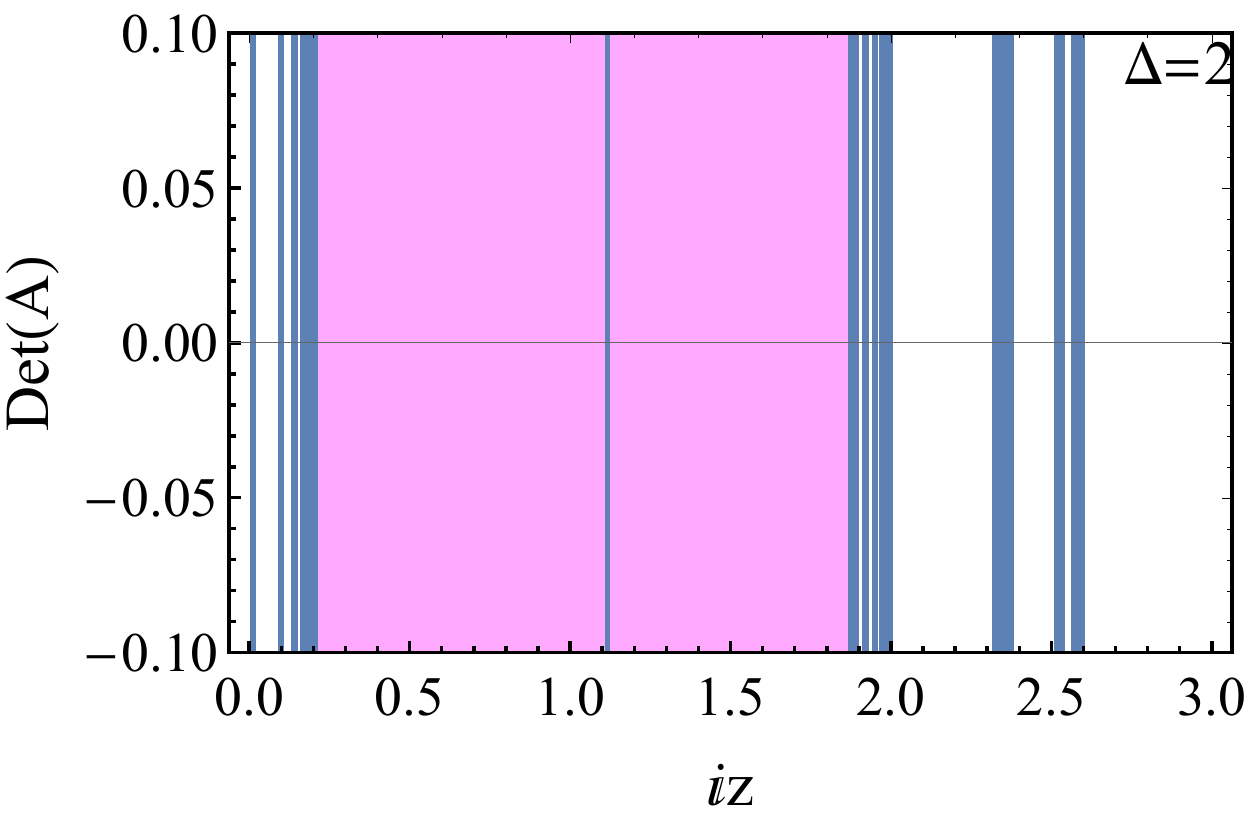}
\includegraphics[width=5.5cm]{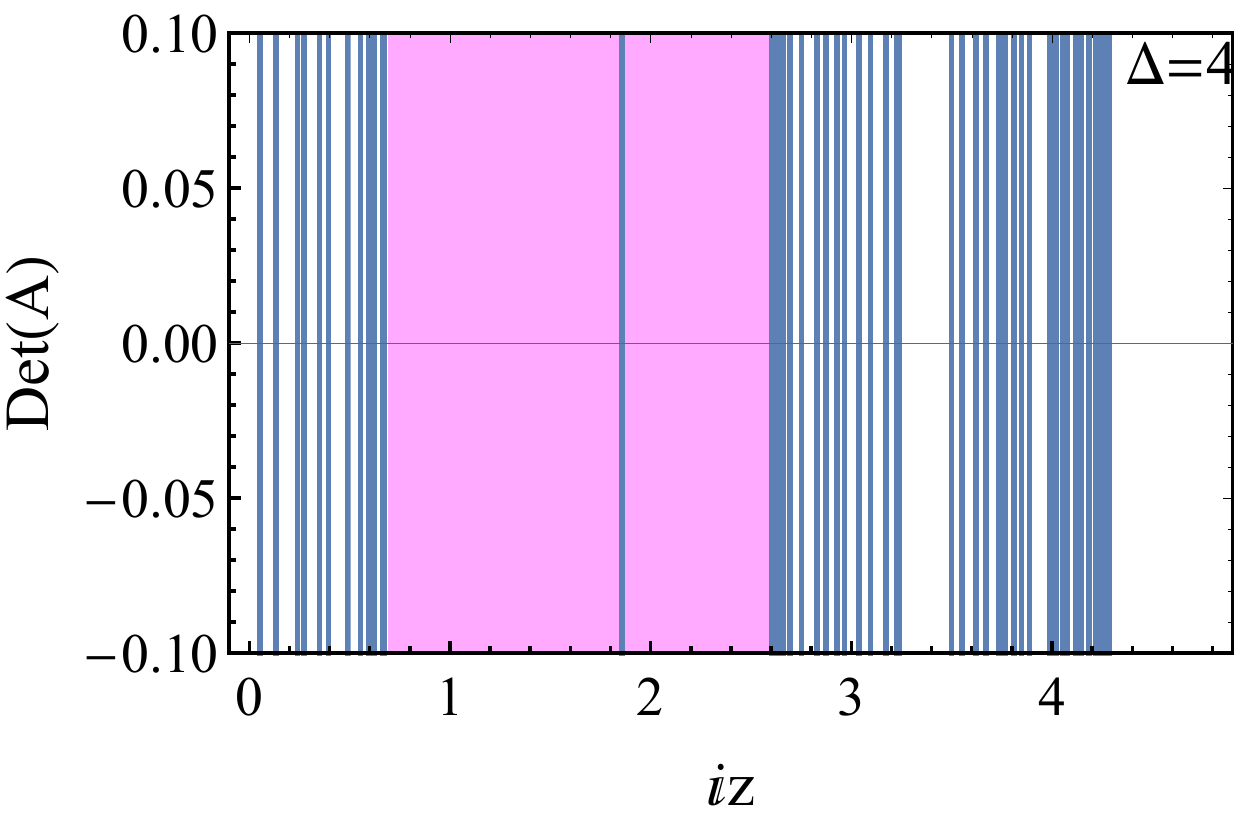}
\caption{(Color online) Plots of  $Det(A)$  for  $\beta=(1+\sqrt{5})/2, \phi=0.4\pi$ when $\Delta=1, 2, 4$ respectively. $N=99$, $s=1, \eta=0.1, \omega_c=10$ are chosen for all plots. The region highlighted by  dark-pink color, denotes the main energy gap  in Fig.\ref{fig:incom-be} }
\label{fig:appendixC1}
\end{figure}

Then the BIC corresponds to a pole of $G_i(z)$ with $\mathbbm{i}z>0$, which can be determined by seeking the solutions to $Det(A)=0$. However because of the involved term $\left(\sum_n\gamma^*_{in}\right)\left(\sum_n\gamma_{jn}\right)$, the result would be different from $\epsilon_i$, as shown in Fig.\ref{fig:appendixC1}. This feature is different from the single qubit case \cite{bic, longhi}, in which  BIC is due to the level resonance. This phenomenon  can be explained by  the level shift, induced by the coupling to a bath. As an example, $Det(A)$ is plotted for positive $\mathbbm{i}z$ for different $\Delta$  in Fig. \ref{fig:appendixC1}.  For these plots, the integral $\Sigma(z)$ is expressed by its principle value. It is clear that a discrete zero point can be found, as shown in Fig. \ref{fig:appendixC1}. Furthermore We find that the positive energy for the discrete zero points are slightly different from the edge mode, which are $0.80462, 1.10176$ and $1.82622$ for $\Delta=1, 2 ,4$ respectively. Besides of the discrete one,  there are  many continuous  zero points, which  construct a band.

\begin{figure}
\center
\includegraphics[width=6cm]{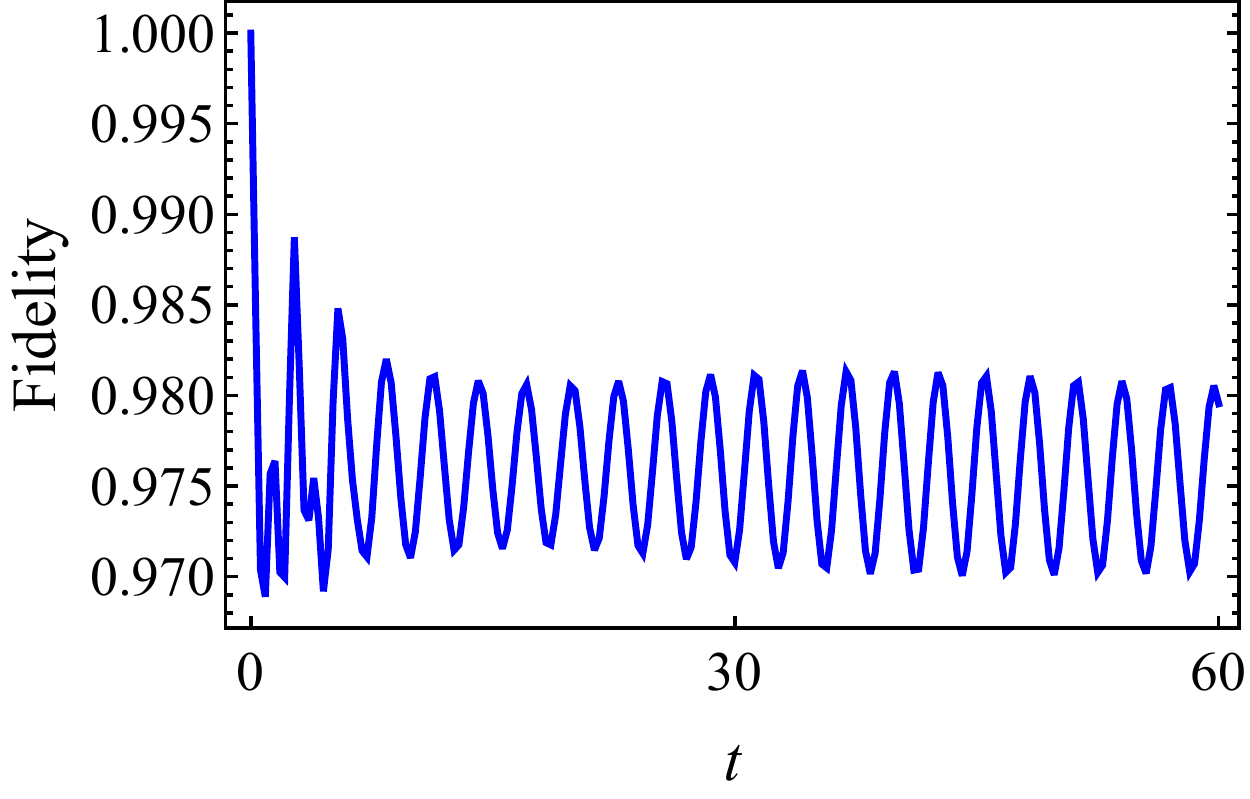}
\caption{(Color online) The evolution of the edge mode as the initial state, occurring for $\phi=0$ when $\beta=1/3, \Delta=2$, is plotted by fidelity. $N=99$, $s=1, \eta=0.1, \omega_c=10$ are chosen for this plot. }
\label{fig:appendixC2}
\end{figure}

Now we will show the correspondence to BIC  for discrete zero point. By inverse Laplace transformation, $\alpha_i(t)$ can be determined. We choose the initial state as the edge state at $\phi=0$ when $\beta=1/3, \Delta=2$ as an exemplification, which corresponds the $47$-th eigenstate in $H_S$. Then one can find by inverse Laplace transformation of $G_{47}(z)$ that the contribution of the discrete zero point at $\mathbbm{i}z=2.30752$ is  $\sim 0.9876 e^{-\mathbbm{i}2.30752 t} \eta^{\dagger}_{47}\ket{0}$. Furthermore we exactly study the evolution dynamics dominated by Eq. (\ref{evolution}) with the edge mode as the initial state. As depicted in Fig. \ref{fig:appendixC2}, the fidelity $\left|\inp{\psi_{edge}}{\psi(t)}\right|^2$ shows a stable oscillation around $0.9876^2\sim 0.975$. Similar observation can be found for the other edge states. Thus we have demonstrated that the discrete poles of $G_i(z)$ characterizes the occurrence of BIC.

\end{document}